\documentclass[fleqn,usenatbib]{mnras}

\usepackage{newtxtext,newtxmath}
\usepackage[T1]{fontenc}

\DeclareRobustCommand{\VAN}[3]{#2}
\let\VANthebibliography\thebibliography
\def\thebibliography{\DeclareRobustCommand{\VAN}[3]{##3}\VANthebibliography}

\usepackage{graphicx}	% Including figure files
\usepackage{amsmath}	% Advanced maths commands
\newcommand{\oiii}{[\ion{O}{iii}]}
\newcommand{\nii}{[\ion{N}{ii}]}
\newcommand{\sii}{[\ion{S}{ii}]}
\newcommand{\oi}{[\ion{O}{i}]}

\title[JWST and ALMA observations of IRAS 20551-4250]{GOALS-JWST: Resolved multi-phase molecular gas in IRAS 20551-4250 using JWST and ALMA}

\author[D. Kakkad et al.]{D. Kakkad,$^{1,2}$\thanks{E-mail: darshankakkad@gmail.com}
Y. Song,$^{3,4}$
T. S.-Y. Lai,$^{5}$
L. Armus,$^{5}$
M. Malkan,$^{6}$
K. L. Larson,$^{7}$
A. S. Evans$^{8,9}$,\newauthor
P. N. Appleton$^{10}$,
L. Barcos-Mu\~{n}oz$^{9,8}$,
M. Bianchin,$^{11}$
T. B\"oker,$^{12}$
T. Bohn,$^{13}$
V. Buiten,$^{14}$
V. Charmandaris,$^{15,16,17}$\newauthor
T. Diaz Santos,$^{15,16}$
H. Inami,$^{18}$
J. Kader,$^{11}$
L. Lenkic,$^{5}$
S. T. Linden,$^{19}$
C. M. Lofaro,$^{16,20}$
G. C. Privon$^{9, 8, 21}$, \newauthor
C. Ricci,$^{22,23}$
M. Sanchez-Garcia,$^{16}$
D. Sanders,$^{24}$
N. Torres-Alba,$^{25}$
V. U,$^{5,11}$
P. van der Werf$^{14}$
\\
$^{1}$Centre for Astrophysics Research, Department of Physics, Astronomy and Mathematics, University of Hertfordshire, Hatfield, AL10 9AB, UK\\
$^{2}$Space Telescope Science Institute, 3700 San Martin Drive, Baltimore, 21210 MD, USA\\
$^{3}$European Southern Observatory, Alonso de Córdova, 3107, Vitacura, Santiago, 763-0355, Chile\\
$^{4}$Joint ALMA Observatory, Alonso de Córdova, 3107, Vitacura, Santiago, 763-0355, Chile\\
$^{5}$IPAC, California Institute of Technology, 1200 East California Boulevard, Pasadena, CA 91125, USA\\
$^{6}$Department of Physics and Astronomy, UCLA, Los Angeles, CA, 90095, USA\\
$^{7}$AURA for the European Space Agency (ESA), Space Telescope Science Institute, 3700 San Martin Drive, Baltimore, MD 21218, USA\\
$^{8}$Department of Astronomy, University of Virginia, 530 McCormick Road, Charlottesville, VA 22903, USA\\
$^{9}$National Radio Astronomy Observatory, 520 Edgemont Road, Charlottesville, VA 22903, USA\\
$^{10}$Caltech/IPAC, MC 314-6, 1200 E. California Blvd., Pasadena, CA 91125, USA\\
$^{11}$Department of Physics and Astronomy, 4129 Frederick Reines Hall, University of California, Irvine, CA 92697, USA\\
$^{12}$European Space Agency, Space Telescope Science Institute, Baltimore, MD 21218, USA\\
$^{13}$Ehime University, Bunkyo-cho 2-5, Matsuyama, Ehime 790-8577, Japan\\
$^{14}$Leiden Observatory, Leiden University, PO Box 9513, 2300 RA Leiden, The Netherlands\\
$^{15}$School of Sciences, European University Cyprus, Diogenes Street, Engomi, 1516 Nicosia, Cyprus\\
$^{16}$Institute of Astrophysics, Foundation for Research and Technology-Hellas (FORTH), Heraklion, 70013, Greece\\
$^{17}$Department of Physics, University of Crete, Heraklion, 71003, Greece\\
$^{18}$Hiroshima Astrophysical Science Center, Hiroshima University, 1-3-1 Kagamiyama, Higashi-Hiroshima, Hiroshima 739-8526, Japan\\
$^{19}$Steward Observatory, University of Arizona, 933 North Cherry Avenue, Tucson, AZ 85721, USA\\
$^{20}$Dipartimento di Fisica e Astronomia, Università di Padova, Vicolo dell’Osservatorio 3, 35122 Padova, Italy\\
$^{21}$Department of Astronomy, University of Florida, P.O. Box 112055, Gainesville, FL 32611, USA\\
$^{22}$Department of Astronomy, University of Geneva, ch. d’Ecogia 16, 1290, Versoix, Switzerland\\
$^{23}$Instituto de Estudios Astrof\'isicos, Facultad de Ingenier\'ia y Ciencias, Universidad Diego Portales, Av. Ej\'ercito Libertador 441, Santiago, Chile\\
$^{24}$Institute for Astronomy, University of Hawaii, 2680 Woodlawn Drive, Honolulu, HI 96822, USA\\
$^{25}$Department of Astronomy, University of Virginia, P.O. Box 400325, Charlottesville, VA 22904, USA\\
}

\date{Accepted XXX. Received YYY; in original form ZZZ}

\pubyear{2024}

\begin{document}
\label{firstpage}
\pagerange{\pageref{firstpage}--\pageref{lastpage}}
\maketitle

% Abstract of the paper
\begin{abstract}
Studying the content and distribution of molecular gas (H$_{2}$) provides key insights into how feedback from Active Galactic Nuclei (AGN) and star formation influences galaxy evolution, since molecular gas is the primary fuel for star formation. Ultra-Luminous Infrared Galaxies (ULIRGs) are ideal candidates to study how AGN and/or starbursts affect the interstellar medium due to their intense AGN and star forming activity. We present spatially-resolved multi-phase molecular gas study of IRAS20551-4250, a nearby ($z=0.0429$) ULIRG, using JWST/MIRI-MRS and ALMA. Mid-infrared diagnostics do not rule out the presence of AGN in IRAS20551-4250. \oiii$\lambda$5007 in VLT/MUSE data reveal ionised gas outflows with $w_{80}^{\rm [OIII]} \sim 790$ km s$^{-1}$ and $\dot{M}_{\rm out}^{\rm[OIII]}<0.01$ M$_{\odot}$ yr$^{-1}$. No outflows are observed in either molecular phases. JWST/MIRI-MRS data reveal several rotational transitions of warm H$_{2}$ (T$\sim500-1400$ K) within the central $\sim4\times4$ kpc$^{2}$ region. Excitation temperature maps suggest that the warm H$_{2}$ is primarily heated by UV radiation from the central source. The CO-based cold molecular component dominates the molecular gas mass, accounting for $>$95\% of the total molecular gas mass.  Warm H$_{2}$ maps show two tidal tails and the velocity centroid maps show disturbed, non-rotational motions and a systematic gradient across the field-of-view, similar to that of ALMA CO-based cold molecular gas and consistent with a late-stage merger. Together, our analysis indicate that the molecular gas composition in IRAS20551-4250 is consistent with ongoing star formation in the host galaxy and the outflows observed in ionised gas phase appear insufficient to expel the molecular gas or quench ongoing star formation.
\end{abstract}

% Select between one and six entries from the list of approved keywords.
% Don't make up new ones.

%%%%%%%%%%%%%%%%%%%%%%%%%%%%%%%%%%%%%%%%%%%%%%%%%%

%%%%%%%%%%%%%%%%% BODY OF PAPER %%%%%%%%%%%%%%%%%%

\section{Introduction} \label{sect1}

Feedback from star formation and accreting supermassive black holes, also called Active Galactic Nuclei (AGN) feedback, have become  primary ingredients of all simulations that model galaxy formation and evolution \citep[e.g.,][]{schaye15, hopkins18, pillepich18, vogelsberger20, schaye23}. Over the past decade, several observations have now confirmed the presence of high velocity outflows (outflow velocity, v$_{\rm out} > $1000 km s$^{-1}$) in low-redshift and high-redshift AGN and star forming galaxies in multiple gas phases, namely neutral, ionised and molecular \citep[e.g.,][]{rupke11, cicone14, genzel14, harrison14, davies20, kakkad20, kakkad22, davies24,  weldon24, vayner24}. These outflows are believed to have the potential to sweep the host galaxies clean of molecular gas and eventually quench star formation \citep[e.g.,][]{piotrowska22}. 

Several efforts on the observational front are underway to understand how these outflows affect the host galaxy properties such as the star formation rate (SFR) using spatially-resolved spectroscopy  \citep[e.g.,][]{cresci15, carniani16, maiolino17, scholtz20, bessiere22, kakkad23a, Dasyra2024}. However, connecting the AGN activity (AGN luminosity or outflow power) with the host's ongoing SFR has not been straightforward, because of the difference in the timescales of the AGN activity, outflow propagation and star formation \citep[e.g.,][]{kennicutt12, schawinski15}. Furthermore, the heterogenous geometry of the galaxies, orientation effects and different AGN types (type-1 and type-2) further makes the calculation of the outflow quantities challenging \citep{Castro2025}. In addition to AGN feedback, feedback from starburst or supernovae may also play a crucial role in driving turbulence \citep[e.g.,][]{koudmani22}. 

A direct method to test the impact of AGN or star formation feedback on host galaxies is by observing the molecular gas itself. The total content of molecular gas may be treated as a reliable indicator of the star forming capacity of galaxies \citep[see reviews by][]{kennicutt_evans12,carilli13, veilleux20, saintonge22}. In the presence of an outflow, it is the gas phase which will be affected first before any impact on star formation is observed. Therefore, any impact on star formation should first be visible as an impact on the overall molecular gas content and/or composition. Molecular gas itself comes in multiple phases and can be broadly divided into three categories depending on the gas temperature. The cold phase (T$<$100 K) is usually traced using transitions in sub-mm spectra (e.g., CO, HCN), the warm molecular gas phase (T = few hundred K) is traced via rotational transitions of the hydrogen molecule in the mid-infrared spectra and the hot phase (1000K$<$T$<3000$K) is traced via ro-vibrational transitions of the hydrogen molecule in the near-infrared spectra \citep[e.g.,][]{maloney88, burton92, kakkad25}.

The majority of molecular gas studies in the literature over the previous decade have focused on the cold gas phase using different transitions of CO. Comparative study of cold molecular gas in mass-matched samples of AGN and star forming galaxies at low and high-redshift have yielded conflicting results, with some showing low molecular gas content in AGN and others showing AGN live in the same or gas rich environment as star forming galaxies \citep[see][]{husemann17, kakkad17, rosario18, circosta21, koss21, bertola24}. These differences may be due to different CO excitation corrections or CO-to-H$_{2}$ conversion factor ($\alpha_{\rm CO}$) between AGN and non-AGN galaxies. Another proposed explanation is that the impact on molecular gas should be more evident in regions closer to the black hole, where a higher abundance of warm and hot H$_{2}$ gas is expected \citep[see][]{u19}. 

\begin{figure*}
\centering
\includegraphics[width=0.96\linewidth]{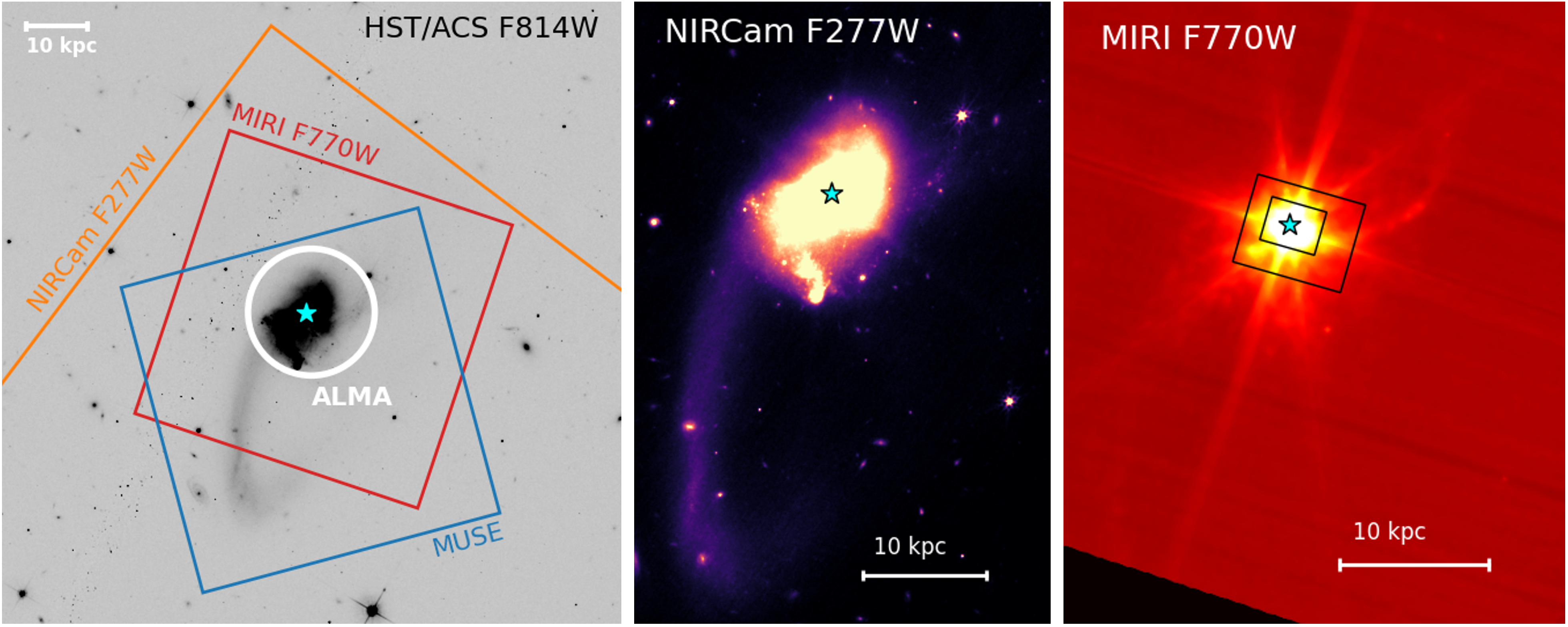}
\includegraphics[width=0.96\linewidth]{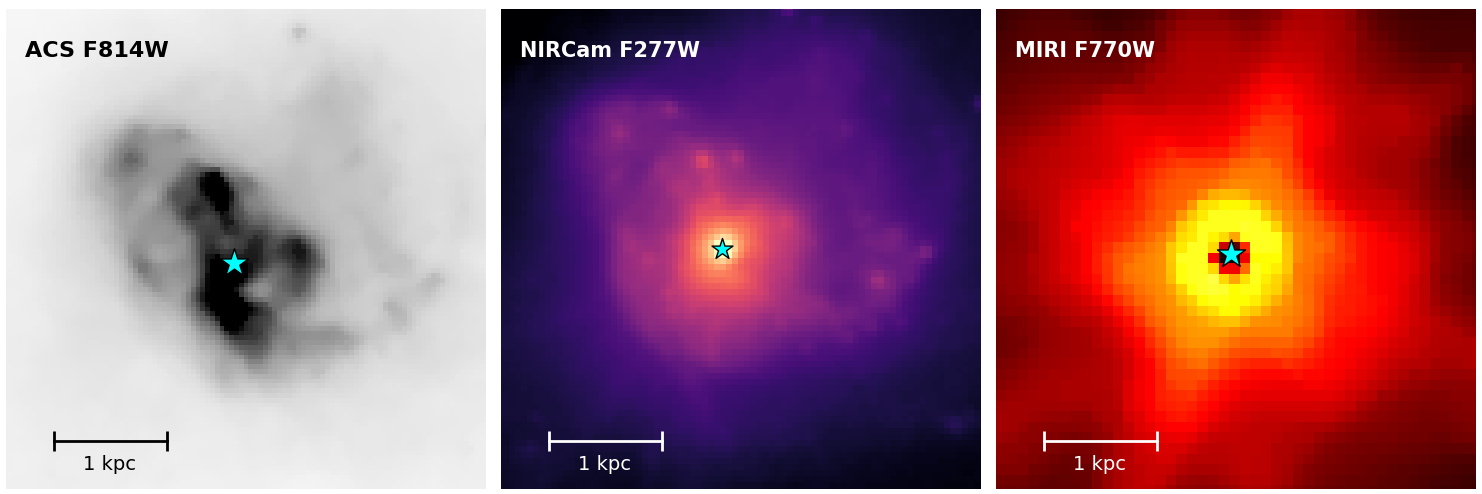}
\caption{The top panels show images from HST/ACS F814W (top left), JWST/NIRCam F277W (top middle) and JWST/MIRI F770W (top right) covering the full range of spatial scales of IR20551. The horizontal bar in all the top panels show the 10 kpc physical scale. The overlaid white circle, red rectangle, orange rectangle and blue rectangle in the top left image shows the ALMA, MIRI, NIRCam and MUSE coverage regions. The black rectangle in the top right panel shows the regions covered by Channels 1 and 4 of the MRS. The bottom panels show the same sequence as the top panels, but zoomed in the central $5\times5$ arcsec$^{2}$ region of the galaxy. The HST and NIRCam images clearly show the presence of extended tidal tail towards the SE direction from the galaxy centre. The MIRI image shows a faint SE tail, but a prominent NW tail, not clearly visible in the HST and NIRCam images.  The HST image does not show a clear point source in the centre, but shows clumpy morphology within the central region, while the NIRCam and MIRI images show a very bright point source at the centre. In fact, the central region in the MIRI image is saturated due to emission from a putative AGN. The horizontal bar shows the 1 kpc physical scale in the bottom panels. The cyan star marks the location of the putative AGN. North is up and East is left in all panels.}
\label{fig:HST_JWST_imgs}
\end{figure*}

Indeed, there has been a parallel effort to get a more detailed, spatially-resolved, view of the molecular gas properties from sub-pc to a few hundred parsec scales, by mapping multi-phase molecular gas in near-infrared, mid-infrared and sub-mm wavelengths using integral field spectroscopy (IFS). For instance, \citet{garcia-burillo21} reported signs of nuclear scale CO-based molecular gas deficits in a sample of low-redshift luminous Seyfert host galaxies, with extreme gas deficits observed in galaxies which host molecular outflows as well. \citet{rosario19} also reported an abundance of ro-vibrational transition, H$_{2}$ 2.12 $\mu$m tracing hot molecular gas, in regions lacking CO-based cold molecular gas \citep[see also][]{feruglio20, davies24, Esparza-Arredondo25}. The emerging picture from these individual multi-phase molecular gas studies is that although the outflows may not remove molecular gas from the host galaxies, they may heat the gas in the circum-nuclear regions via shocks or X-ray radiation.
%to higher temperatures. 

In this regard, Luminous infrared galaxies (LIRGs) and ultra-luminous infrared galaxies (ULIRGs) are particularly interesting targets to study the impact of feedback from starbursts and/or AGN on the hosts' molecular gas. ULIRGs are often a result of mergers and the exchange of gas during the merger results in an intense starburst and at the same time, fuels gas and dust onto the central supermassive black hole. The primary radiation field in ULIRGs is often reprocessed by dust, which results in very high infrared luminosities. The intense radiation, combined with a gas rich environment, means ULIRGs are prime sites for hosting fast outflows \citep[e.g.,][]{pereira-santaella18, lamperti22}. As a result, shocked molecular gas or molecular gas heated by X-rays is expected to be common in these galaxies \citep[see][]{Sugai1997,u19}.

In this paper, we present multi-phase gas properties of a nearby ULIRG, IRAS 20551-4250 (also named ESO 286-19 or ESO 286-IG019, RA = 20:58:26.82, DEC = $-$42:38:59.41, $z$ = 0.0429). Hereafter, we simply refer to this system as IR20551. IR20551 is a part of the Great Observatories All-sky LIRG Survey \citep[GOALS, see][]{armus09}. The GOALS survey consists of a sample of bright ($S_{60} > 5.24$ Jy) LIRGs, selected from the IRAS revised bright galaxy sample \citep[see][]{sanders03}. IR20551 was first discovered within the photometric and spectroscopic ESO/Uppsala programme, as reported in \citet{west78}. Its exceptionally bright infrared nature was confirmed in \citet{johansson91}, which was attributed to warm dust heated by young stars. Subsequently, IR20551 has been targeted by several multi-wavelength campaigns to unveil the true nature of the host galaxy and its nucleus \citep[e.g.,][]{genzel98, misaki99, franceschini03, farrah07, haan11, sani12, stierwalt13, stierwalt14, imanishi16}. These multi-wavelength observations have suggested that IR20551 is an interacting system in a fairly advanced stage of merger with a disturbed core and a prominent tidal tail (Figure \ref{fig:HST_JWST_imgs}). It is very bright in the infrared with $L_{\rm IR} = 1.15\times 10^{12}$ L$_{\odot}$ \citep{armus09}, placing the target in the ULIRG category. 

IR20551 is a composite system, which shows possible signatures of both AGN activity as well as a starburst, depending on the waveband of observation. For example, at optical wavelengths, this target is classified as an \ion{H}{ii} region \citep{johansson91} and mid-infrared observations indicate both starburst and possible AGN contribution \citep[e.g.,][]{genzel98}. The AGN nature of IR20551 is also indicated based on its hard X-ray luminosity, log $L_{\rm X(2-10 ~keV)} = 42.8$ erg s$^{-1}$ \citep[close to the threshold of AGN classification log $L_{\rm X(2-10 ~keV)} > 42.5$ erg s$^{-1}$][]{padovani17, hickox18, Saade2022} and an X-ray column density, $N_{\rm H} = 8 \times 10^{23}$ cm$^{-2}$ \citep[e.g.,][]{franceschini03}. A recent joint NuSTAR$+$XMM$+$Chandra analysis of this source did not rule out the presence of an obscured AGN in this source \citep[see][]{yamada21}. The implied X-ray column density suggests that the AGN, if confirmed, is heavily obscured. A quantitative estimate for the contribution of the AGN to infrared emission was reported by \citet{farrah07}, who suggest that due to the heavy obscuration, typical AGN tracers such as [\ion{Ne}{v}] and [\ion{O}{iv}] \citep{Spinoglio2022} are absent in its mid-IR spectra. Atacama Large Millimetre-submm Array (ALMA) observations of IR20551 over the last decade have also confirmed the presence of dense molecular gas, as traced by species such as HCN or HCO$+$, suggesting an abundant molecular gas supply able to sustain star formation or feed AGN in the centre \citep[e.g.,][]{imanishi16}.

In this paper, we present detailed multi-wavelength imaging and IFS observations of IR20551. Specifically, we will present the mid-infrared spectral properties using the Medium Resolution Spectrograph (MRS) of the Mid-InfraRed Intrument (MIRI) aboard JWST and CO(2-1) spectra using ALMA. We compliment these data with archival Multi-Unit Spectroscopic Explorer (MUSE) data at the Very Large Telescope (VLT) and the Advanced Camera for Surveys (ACS) aboard Hubble Space Telescope (HST). The aims of this paper is to investigate the true nature of the nuclear regions in this galaxy, characterise cold and warm molecular gas properties and infer if the molecular gas observations provide any evidence of feedback from starburst or AGN, as is often expected in ULIRGs. This paper is organised as follows: In Section \ref{sect2}, we present the details of the observing datasets used from JWST, ALMA and the VLT. Following this, in Sect. \ref{sect3}, we present our analysis methods for each dataset and the results alongside them and in Sects. \ref{sect4} and \ref{sect5}, we discuss the implications of these results and the conclusions of this work, respectively. Throughout this paper, we use the following cosmology: $H_{0}$ = 70 km s$^{-1}$, $\Omega_{\rm M}$ = 0.3 and $\Omega_{\Lambda}$ = 0.7. Unless mentioned otherwise, North is up and East is to left in all the maps presented in this paper. 

\section{Observations and data reduction} \label{sect2}

In this section, we describe the observations and data reduction of the multi-wavelength data presented in this paper.

\subsection{JWST/MIRI-MRS} \label{sect2.1}

IR20551 was observed on 16 October 2023 using MRS \citep{wells15}, as a part of the Cycle-2 General Observer (GO) programme PID3368 (co-PIs: Armus \& Evans). No target acquisition was performed, as the pointing accuracy of 0.1 arcsec was sufficient to centre the target in the MRS field-of-view. The observation consisted of a single IFU pointing, centred on the infrared-bright nucleus. We used a 4-point dither pattern, which improved the spatial sampling and resolution of the final drizzled data cube \citep[see][]{law23}. Separate background exposures were taken along with the on-source observations as a non-interruptible sequence to subtract the infrared background emission, especially in Channels 3 and 4. During the background observations, simultaneous MIRI imaging of the target was performed in filters F560W, F777W and F1500W. Given the very bright nucleus, the BRIGHTSKY subarray was used to avoid saturation. 

The MRS observations covered all four channels and their sub-bands (SHORT, MEDIUM and LONG), resulting in a contiguous wavelength coverage between 4.9--27.9 $\mu$m ($\sim$4.7--26.7 $\mu$m in the rest-frame) with a resolving power $\sim$1300-3700 (highest resolving power in the short-wavelength channels). The FoV varies between $\sim 3.2\times 3.7$ -- $6.6\times 7.7$ arcsec$^{2}$ and PSF between 0.25--1.0 arcsec from Channels 1--4. The observations employed a FASTR1 readout mode for a single exposure with 80 groups per integration and one integration per exposure. The total exposure time was 888s per channel. 

We used the standard MRS data reduction procedure using the JWST pipeline \citep[version: 1.15.1, see][]{labiano16} using the Calibration Reference Data System (CRDS) context \texttt{jwst\_1256.pmap}. Wherever necessary, we also employed additional customised steps and fine-tuned pipeline parameters during the data reduction to improve data quality which are briefly summarised here. There are three stages to the MRS data reduction pipeline: The first stage performs detector-level corrections, where we turned on the \texttt{find\_showers} keyword and kept the rest of the default pipeline parameters unchanged. The \texttt{find\_showers} keyword helps identify and correct for Cosmic Ray (CR) showers. Some warm pixels still remain unidentified in this stage. To remove these, we derived the median of all slope images (\texttt{*rate.fits} files) obtained in the output of stage-1 and applied a standard sigma-clipping algorithm to flag the outliers as \texttt{DO\_NOT\_USE} in the bad pixel mask. 

The files with the updated bad pixel masks were then processed during the second stage of the pipeline, which performs instrument calibrations such as flux and wavelength calibration and flat-field correction. This stage also performs additional corrections to residual fringes in the individual channels by setting the parameter \texttt{skip\_residual\_fringe} to \texttt{FALSE}. The background subtraction was performed using the dedicated background exposures. The output of this second stage of the pipeline is a set of fully calibrated individual exposures (\texttt{*cal.fits} files). 

These fully calibrated datasets are then fed to the third stage of the pipeline, which combines exposures across different channels and sub-bands into a final drizzled data cube. The parameters for the third stage of the pipeline were mostly left unchanged, except changing the outlier threshold at 99\%. During the first stage of the pipeline, a pixel atop a bright trace of a source might get flagged as \texttt{DO\_NOT\_USE}. During the cube building in stage-3, values then have to be interpolated from other dither positions that may have a different centre, resulting in absorption-like drop outs in the spectrum. To avoid such effects, we turned on the \texttt{pixel\_replace} step in the third stage of the pipeline (in the earlier versions of the pipeline, this parameter used to be in the second stage).

The final data products consist of 12 data cubes, corresponding to each channel (1, 2, 3 and 4) and the respective sub-bands (SHORT, MEDIUM and LONG configurations), providing a continuous wavelength coverage between 4.9--27.9 $\mu$m.

\begin{figure*}
\centering
\includegraphics[width=0.96\textwidth]{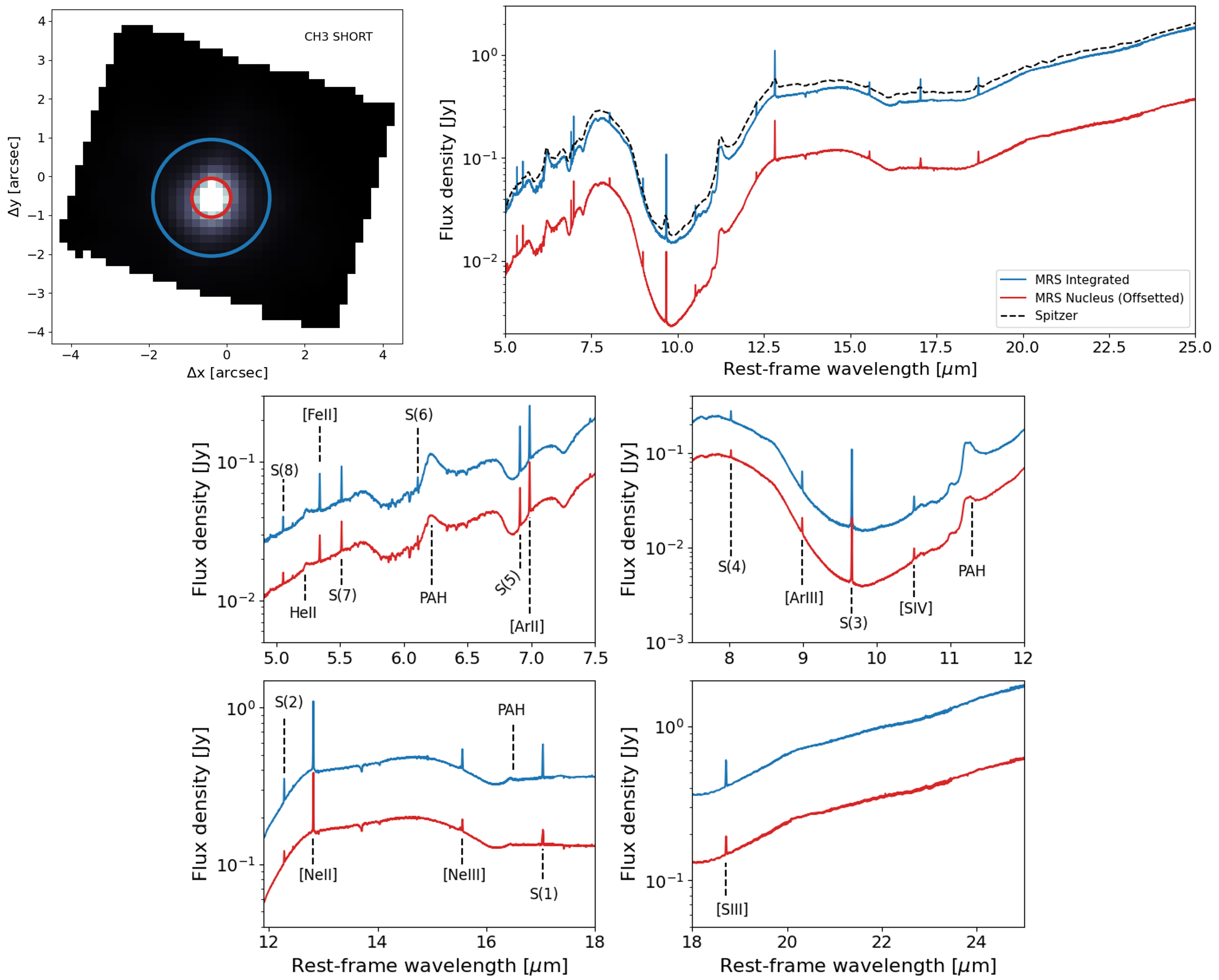}
\caption{Top left panel shows the median MRS image of IR20551 in Channel-3. The red and blue circles show the aperture of extraction for nuclear (R = 0.5 arcsec) and integrated spectra (R = 1.5 arcsec), respectively. The top right panel shows the full MRS spectra from all four channels from the nuclear (red) and integrated (blue) apertures. The nuclear aperture is offset by an arbitrary factor for better visualisation. Archival Spitzer spectra of IR20551 is shown as black dashed line, which showcases the improvement in the quality of spectra with JWST/MIRI-MRS. The MRS spectra of IR20551 shows a combination of narrow lines, PAH emission and silicate absorption features at $\sim 6.2 \mu$m and $\sim 9.7 \mu$m. The bottom four panels show the same spectra as the top right panel, but zoomed in on four different regions of the MRS spectra and the strong emission line features labelled. Rotational transitions of H$_{2}$ (S(1)--S(8)) are detected, along with PAH and low-ionisation lines such as [\ion{Ne}{ii}]. High ionisation emission lines such as [\ion{Ne}{v}]14.322 are not detected or are weak (see also Fig. \ref{fig:Ne_specs}).}
\label{fig:nuclear_integrated_spec}
\end{figure*}

\subsection{Archival ALMA and VLT/MUSE Observations} \label{sect2.2}

The JWST/MIRI-MRS observations described in Sect. \ref{sect2.1} trace the warm phase ($\sim$100-1000 K) of the molecular gas. However, molecular gas also exists in the cold gas phase, observable at mm/sub-mm wavelengths, and the different sources of ionisation (star formation/AGN) can be inferred from diagnostics in optical spectra \citep[e.g., BPT diagrams:][]{veilleux87}. Therefore, we complemented our JWST/MIRI-MRS observations with resolved sub-mm observations from ALMA and archival optical IFS observations from VLT/MUSE. 

We used Phase-3 ESO data release portal to retrieve the fully processed MUSE data cubes. IR20551 was observed with MUSE on 10 September 2015 as a part of a GTO programme (ID 095.B-0049, PI: Soto). The average seeing during the observations was $\sim$1.5 arcsec, as recorded in the observatory logbooks. The spatial resolution of MUSE observation is $\sim$1.3 arcsec, derived from the FWHM of the isolated stars in the MUSE field-of-view. The IR20551 MUSE data cube has a field-of-view of $\sim 1.2\times 1.2$ arcmin$^{2}$ with a spaxel sampling of $0.2 \times 0.2$ arcsec$^{2}$. The spectral axis provides a contiguous wavelength coverage in the optical from 0.465-0.93 $\mu$m, with a spectral resolving power ranging from 2000 at the shortest wavelengths to 4000 at the long wavelength end. 

Archival ALMA CO(2-1) observations of IR20551 were obtained from the programme 2022.1.01262.S (PI: J. Ueda). Observations were done in two separate configurations covering baselines up to $\sim 2.6$ km (TM1) and $\sim 0.6$ km  (TM2), with array angular resolution of $\sim 0.06$ arcsec and $\sim 0.26$ arcsec, and maximum recoverable scales of $\sim 1.0$ arcsec and $\sim 3.7$ arcsec, respectively. We recreate the calibrated Measurement Sets (MSs) using CASA pipeline version 6.5.4-9-pipeline-2023.1.0.124 \citep{casateam}. Each MS was trimmed down to contain only the UV visibilities from the spectral window containing the CO(2-1) line, afterwards the 0th order polynomial fit to the continuum emission in the spectral window was then subtracted using the \texttt{uvcontsub} task. We perform imaging of the CO(2-1) line in 4.5 km s$^{-1}$ channels combining data from both MSs with \texttt{tclean},  adopting Briggs weighting \citep{briggs} with robust of 2.0, \texttt{auto-multithresh} masking \citep{automasking}, and the \texttt{multi-scale} deconvolver \citep{mulitscale} to model emission on scales of 0.1, 0.5, 1.0, and 3.0 arcseconds. We use the interactive mode to visually inspect and manually edit the cleaning mask generated at each cycle as needed, until the cycle threshold reaches $\sim 3\sigma_{\rm RMS}$, where $\sigma_{\rm RMS} = 0.44$ mJy/beam is the per channel root-mean-square value measured within an emission free region in the dirty cube (i.e. before deconvolution). We then repeat the above process but with the \texttt{hogbom} deconvolver \citep{hogbom} to clean more deeply until the residual level reaches $\sigma_{\rm RMS}$ \citep[see][]{leroy21}. The final CO(2-1) cube has a synthesized beam of $0.12\times0.10$ arcsec$^{2}$. Lastly, we smooth the primary beam-corrected CO cube to a Gaussian beam with FWHM of $0.8$ arcsec and re-grid the cube to match the worst spatial resolution of the MRS data in Channel-4. 

\section{Data analysis} \label{sect3}

In this section, we describe the spectral analysis tools to model the continuum and emission lines in MRS, ALMA and MUSE data. 

\subsection{Modelling warm molecular gas transitions in MRS data} \label{sect3.1}

The MRS data span a large range of wavelength from 4.7 $\mu$m -- 26.7 $\mu$m across all four channels, each of which has a different spaxel scale, spectral resolution and field-of-view. Our analysis is broken down into two steps: We first extract the spectra from each MRS cube (Channels 1-4: Short, medium and long) using two circular apertures with radii 0.5 and 1.5 arcsec (corresponding to physical scales of $\approx$0.4 and 1.3 kpc, respectively), which we call nuclear and integrated spectra. The continuum levels of the extracted spectra from different channels/sub-channels do not perfectly align with each other. To correct this, we stitched the spectra together by adjusting the continuum levels across channels, using the lower channel as the reference. Using these aperture spectra, we get a first-order idea of the characteristics of the mid-infrared spectrum of IR20551, such as the presence of Polycyclic Aromatic Hydrocarbon (PAH) emission, silicate absorption features and relevant emission lines such as the H$_{2}$ rotational transitions. We also compare the nuclear and integrated spectra with archival Spitzer/IRS spectra and demonstrate the incredible improvement in depth and spectral resolution of MRS compared to these archival spectra. We then focus on fitting the individual warm molecular hydrogen rotational transitions with a Gaussian model in the nuclear and integrated spectra, followed by a spatially-resolved pixel-by-pixel analysis of these lines.

\begin{figure*}
\centering
\includegraphics[width=0.48\textwidth]{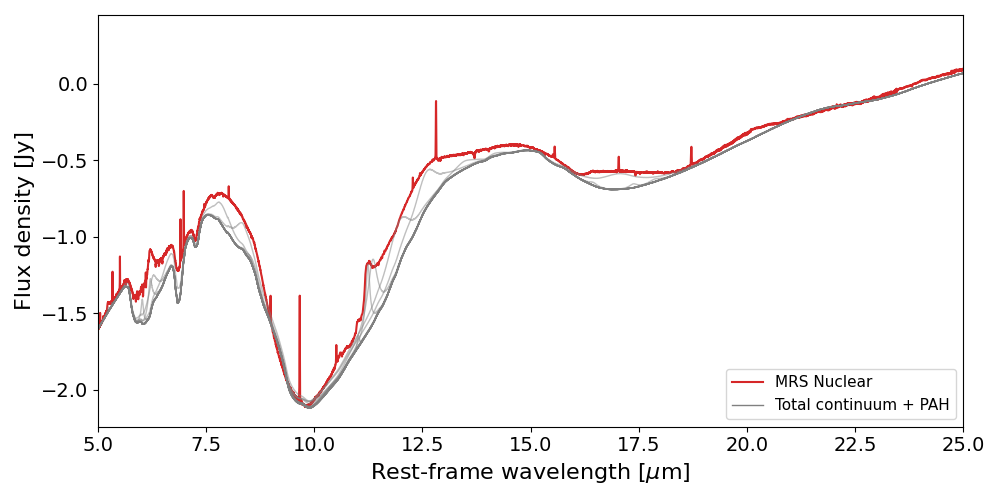}
\includegraphics[width=0.48\textwidth]{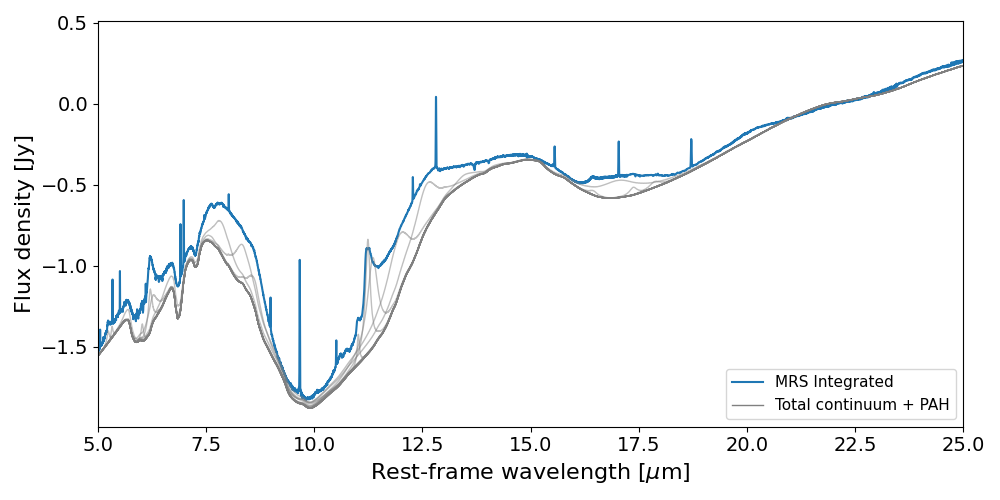}
\caption{The nuclear (left) and integrated (right) MRS spectra of IR20551, with the total continuum model and PAH emission line fits from \texttt{CAFE} overlaid. The individual components from AGN, starburst, hot, warm and cold dust etc are not shown due to high degree of degeneracy between these different components. The \texttt{CAFE} model reproduces the overall mid-infrared continuum well, along with the PAH emission and silicate absorption features.}
\label{fig:CAFE_fits}
\end{figure*}

\begin{figure*}
\centering
\includegraphics[width=0.95\textwidth]{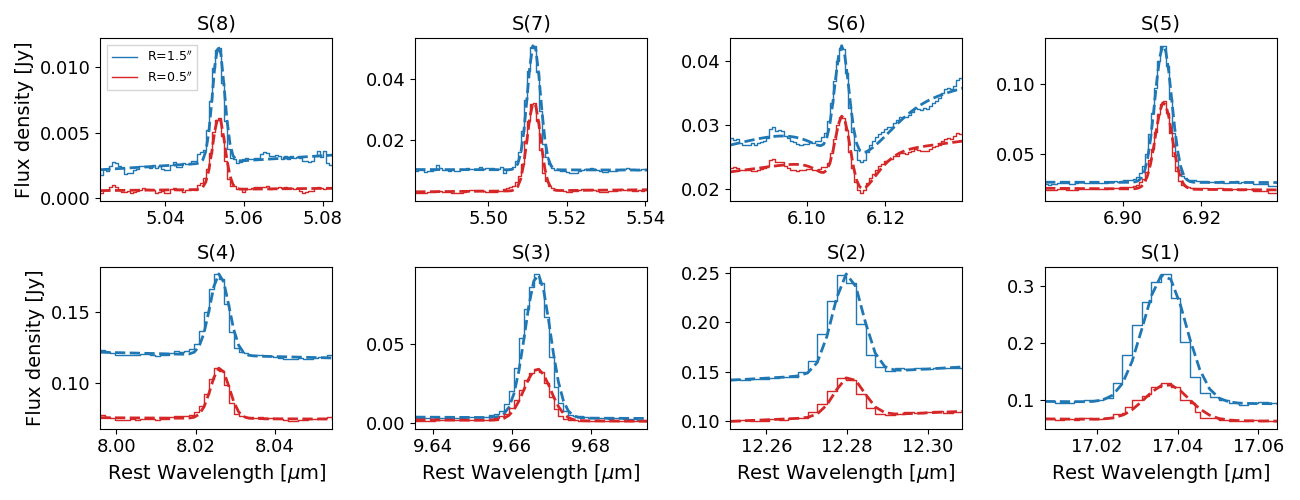}
\caption{Rotational hydrogen lines (S(1) -- S(8)) from the nuclear (solid red) and integrated (solid blue) spectra and respective single Gaussian fits (dashed lines). The rotational transitions do not show any asymmetry and are reproduced using a single Gaussian function, suggesting that there is no evidence for fast moving warm molecular gas associated with an outflow. S(6) line is affected by an absorption feature towards the right of the profile from the instrument, which is well fit with an inverted Gaussian component.}
\label{fig:h2_specs}
\end{figure*}

\begin{table*}
\centering
\begin{tabular}{lccccccc} 
\hline
& & \multicolumn{3}{c}{R = 0.5 arcsec} & \multicolumn{3}{c}{R = 1.5 arcsec}\\
Line & $\lambda_{\rm rest}$ & $\Delta v^{\rm offset}_{\rm obs}$ & FWHM & Flux & $\Delta v^{\rm offset}_{\rm obs}$ & FWHM & Flux \\
& $\mu$m & km/s & km/s & 10$^{-14}$ erg/s/cm$^{2}$ & $\mu$m & km/s & 10$^{-14}$ erg/s/cm$^{2}$\\
(1) & (2) & (3) & (4) & (5) & (6) & (7) & (8)\\
\hline\hline
0-0 S(8) & 5.053 & -12 & 194$\pm$4 & 1.14$\pm$0.02 & -12 & 186$\pm$11 & 1.69$\pm$0.07\\
0-0 S(7) & 5.511 & -30 & 189$\pm$2 & 4.87$\pm$0.05 & -30 & 193$\pm$2 & 6.93$\pm$0.07\\
0-0 S(6) & 6.110 & 52 & 203$\pm$15 & 2.22$\pm$0.24 & 52 & 186$\pm$10 & 2.70$\pm$0.14 \\
0-0 S(5) & 6.910 & -23 & 187$\pm$3 & 8.69$\pm$0.10 & 16 & 190$\pm$3 & 13.13$\pm$0.12\\
0-0 S(4) & 8.025 & -26 & 185$\pm$4 & 4.15$\pm$0.08 & -26 & 188$\pm$4 & 6.63$\pm$0.11\\
0-0 S(3) & 9.665 & -11 & 228$\pm$3 & 10.32$\pm$0.12 & -11 & 212$\pm$3 & 26.91$\pm$0.25\\
0-0 S(2) & 12.279 & -28 & 182$\pm$5 & 3.54$\pm$0.06 & -28 & 189$\pm$2 & 10.07$\pm$0.13\\
0-0 S(1) & 17.035 & -37 & 217$\pm$6 & 4.91$\pm$0.14 & -37 & 221$\pm$5 & 17.67$\pm$0.41\\
\hline
\end{tabular}
\caption{Basic properties of the emission lines measured in the MRS spectra of IR20551 from apertures of radii, R = 0.5 and 1.5 arcsec. (1): Emission line feature, (2): The rest-frame wavelength of the line in $\mu$m, (3), (4) \& (5): observed velocity offset of the emission line from the expected location based on its redshift (determined from the location of \oiii$\lambda$5007 line in the optical MUSE spectra), line width (FWHM, corrected for the spectral resolution) and line flux within the 0.5 arcsec aperture, (6), (7) \& (8): same as (3), (4) and (5), respectively, but within the 1.5 arcsec aperture. The quoted errors indicate 1$\sigma$ uncertainty.}
\label{tab:h2_lines}
\end{table*}

\begin{figure*}
\centering
\includegraphics[width=0.8\textwidth]{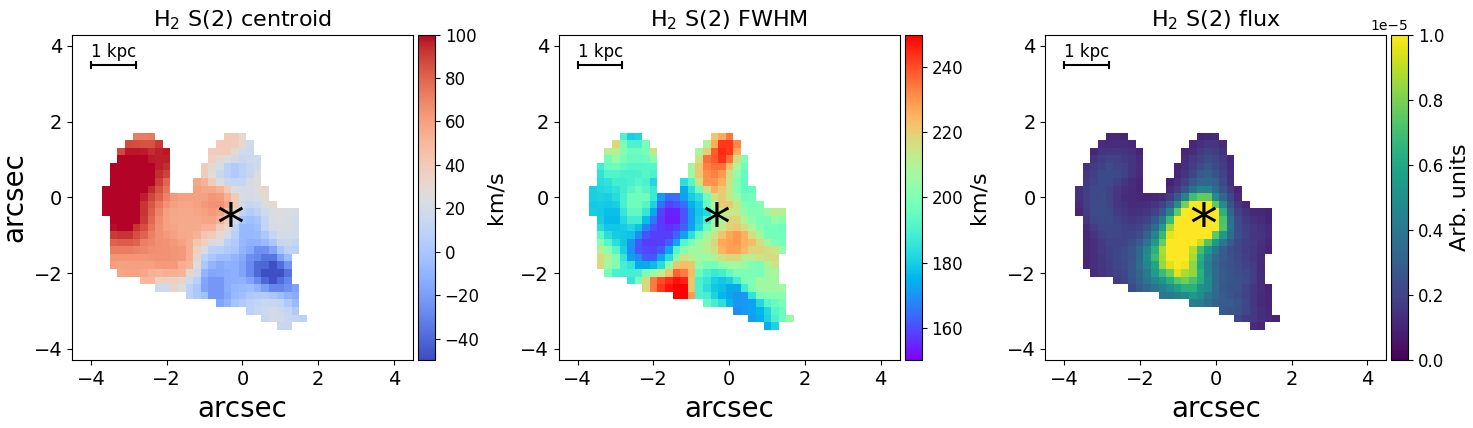}
\includegraphics[width=0.8\textwidth]{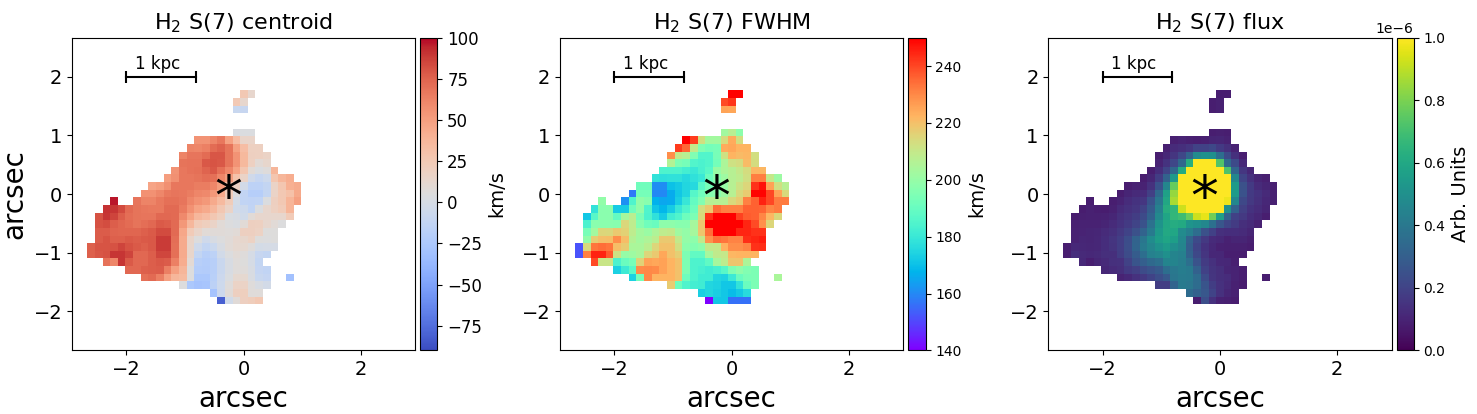}
\caption{Maps showing the variation in the centroid (left panels), width (FWHM, middle panels) and flux (right panel) of the H$_{2}$ rotational transitions, S(2) (top panels) and S(7) (bottom panels) as examples. The maps showcase the presence of two tidal tails in the central region of IR20551 - one towards the NE and other towards the SE. S(7), a higher excitation line, is covered in the lower channels of MRS, which have a smaller field-of-view. Therefore, the tidal tail towards the NE of the nucleus, seen in S(2), is not visible. The black star in all plots mark the location of the nucleus (putative AGN). See Sect. \ref{sect4.2} for more details about these maps.}
\label{fig:h2_maps}
\end{figure*}

The spectra extracted from the nuclear (R = 0.5 arcsec) and integrated (R = 1.5 arcsec) apertures are shown in Figure \ref{fig:nuclear_integrated_spec}. To model the continuum emission in each spectra, we use a modified version of the \texttt{CAFE} software\footnote{https://github.com/GOALS-survey/CAFE} \citep[see][]{marshall07, lai22, armus23, cafe_software} that was originally adopted for Spitzer/IRS data and is currently being updated for JWST. \texttt{CAFE} simultaneously models the PAH emission, dust continuum, silicate absorption, and the narrow fine structure atomic and molecular gas emission lines. As the main focus of our paper is to analyse the warm molecular gas transitions, we model the overall continuum emission, without a detailed decomposition into individual contributions from the AGN, starburst or different dust components. As there are many components, such continuum decompositions are also prone to degeneracies. However, based on the saturated central source in F770W image in Fig. \ref{fig:HST_JWST_imgs}, it is likely that the mid-infrared continuum is dominated by the central source (a putative AGN) in IR20551. Figure \ref{fig:CAFE_fits} shows the mid-infrared continuum and PAH components in the nuclear and integrated MRS spectra of IR20551. 

\begin{figure}
\centering
\includegraphics[width=0.8\linewidth]{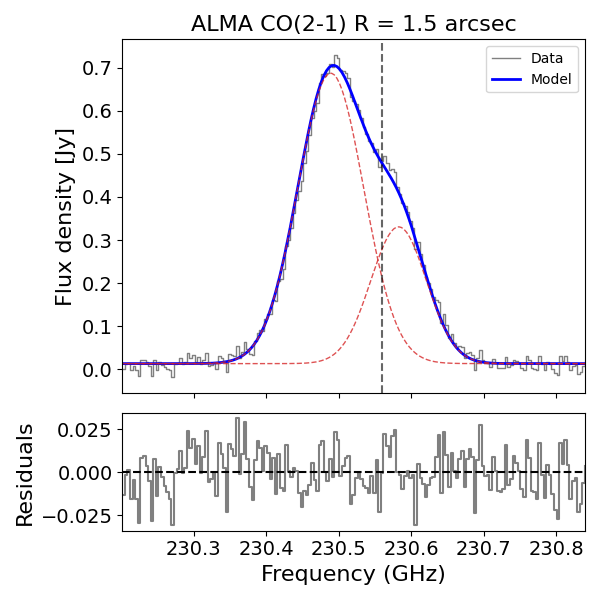}
\caption{The grey curve in the top panel shows the extracted CO(2-1) spectra from the ALMA data cube (radius of aperture, R = 1.5 arcsec) and the blue curve overlaid shows the total multi-Gaussian model. The dashed red curves show the individual Gaussians. The vertical line shows the expected location of CO(2-1) line based on the redshift of IR20551. The bottom panel shows the residuals from the fit. The $w_{80}$ parameter for the total Gaussian fit is $\sim$200 km s$^{-1}$, which does not suggest the presence of a fast moving cold molecular gas outflow in IR20551.}
\label{fig:CO_intspec}
\end{figure}

\begin{figure*}
\centering
\includegraphics[width=0.8\textwidth]{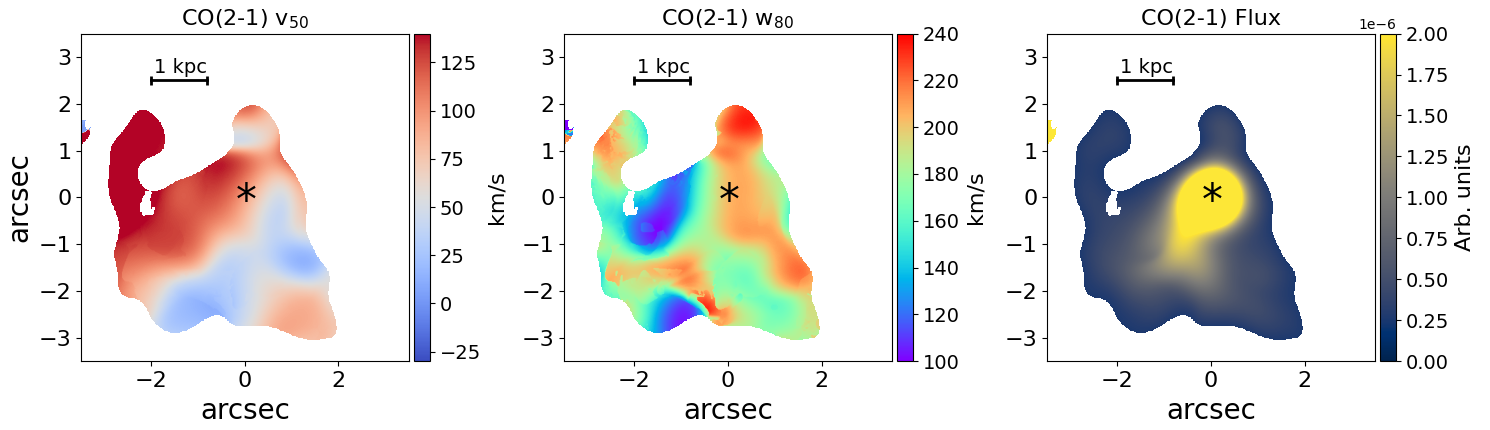}
\caption{CO(2-1) maps of IR20551: $v_{50}$ which is equivalent of a centroid map is shown in the left panel, the middle panel shows the $w_{80}$ map (equivalent of a FWHM map) and the right panel shows the flux map. Non-parametric measurements are shown here due to the double Gaussian fitting in CO(2-1) lines. The black star in the centre of the images mark the putative AGN location. The centroid, width and morphology mirror the warm molecular gas transitions shown in Fig. \ref{fig:h2_maps}, suggesting that the two molecular gas phases show similar morphological and kinematic properties.}
\label{fig:CO_maps}
\end{figure*}

After fitting the underlying mid-infrared continuum, we focus on modelling the rotational transitions of the H$_{2}$ molecule in the nuclear and integrated spectra. The observed spectrum contains S(1) -- S(8) H$_{2}$ transitions, with their expected wavelengths (based on the redshift of IR20551) listed in Table \ref{tab:h2_lines}. Although the \texttt{CAFE} software models these individual H$_{2}$ and other ionised emission lines in the spectra, there are significant residuals at the location of these emission lines. Therefore, we only use the overall continuum models from \texttt{CAFE}, and use a custom script to model the S(1) --  S(8) H$_{2}$ lines after subtracting the overall continuum. Each H$_{2}$ line is fitted with a single Gaussian function around spectral windows within $\pm 0.03\mu$m of the respective line centres. We did not add additional Gaussians, based on minimal changes in the Bayesian Information Criterion \citep[BIC, see][]{gideon78} values between single and multiple Gaussian models. The H$_{2}$ emission lines largely exhibit narrow profiles with no apparent red or blue wings. To estimate the uncertainties on the derived parameters, we generated 100 mock spectra by adding rms noise to the modeled spectrum and repeated the line-fitting procedure for each realization. The standard deviation of the resulting parameter distributions was then adopted as the error estimate. The H$_{2}$ emission line profiles and the corresponding Gaussian fits for the nuclear and integrated aperture spectra are shown in Fig. \ref{fig:h2_specs}. The parameters of the Gaussians fits (line centroid, FWHM and flux) are reported in Table \ref{tab:h2_lines}. The reported line widths (FWHM) in Table \ref{tab:h2_lines} are corrected for the instrumental LSF. 

We also fit each H$_{2}$ line spaxel-by-spaxel with a single Gaussian function and using the parameters from the integrated line fitting as a prior. For these spaxel-by-spaxel fits, we allow the line widths and the line centroids to vary in order to reproduce the variation in the H$_{2}$ line profiles due to galaxy rotation and/or turbulence. We note that  IR20551 is a late-stage merger and therefore, we expect turbulence in gas kinematics, especially in the central regions. The output of the spaxel-by-spaxel fitting consists of the line centroid, width and flux map for each H$_{2}$ transition. We show the maps for S(2) and S(7) transitions as examples in Fig. \ref{fig:h2_maps}. The rest of the H$_{2}$ transitions show similar morphological and kinematic structure as S(2) and S(7). 

\subsection{Resolved CO(2-1) characterisation with ALMA} \label{sect3.2}

\begin{figure*}
\centering
\includegraphics[width=0.99\linewidth]{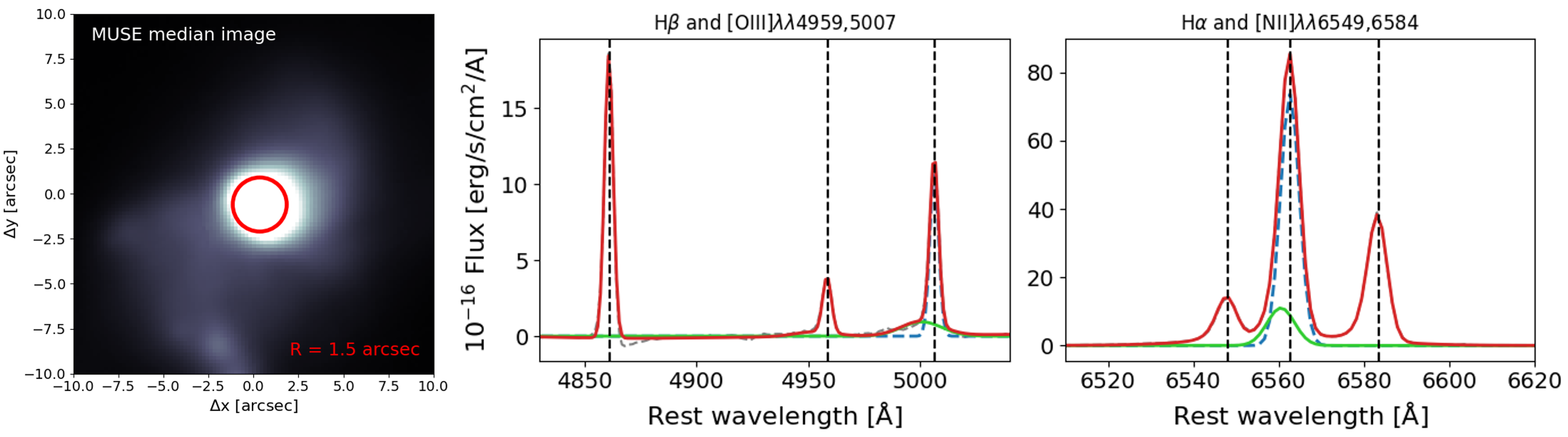}
\includegraphics[width=0.99\linewidth]{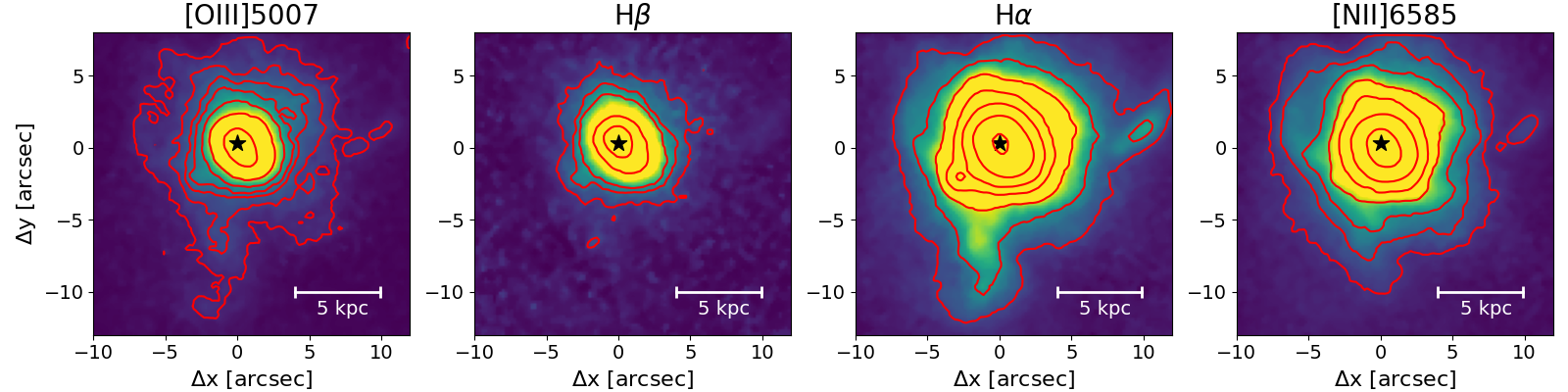}
\caption{Top left panel shows the median MUSE image of IR20551 and the red circle shows the aperture of optical spectral extraction (R = 1.5 arcsec), similar to the aperture radii used for MRS integrated and ALMA CO(2-1) data. Top middle and right panels show the extracted MUSE spectra with emission line model fits, zoomed in on H$\beta$, \oiii$\lambda\lambda$4959, 5007 (middle panel), H$\alpha$ and \nii$\lambda\lambda$6549, 6585 (right panel). The background grey curve in the top middle and top right panels is the original data, the red curve shows the overall fitting, the dashed blue curve shows the first Gaussian component (narrow) and the solid green line shows the second Gaussian component (broad). The bottom panels show the flux maps of these emission lines with red contours added for visualisation of the most extended low surface-brightness line emission. H$\alpha$ emission shows a second clump to the SE of the AGN location, which is a star- forming region confirmed via BPT map in Fig. \ref{fig:bpt}.}
\label{fig:muse_fluxmaps}
\end{figure*}

\begin{figure*}
\centering
\includegraphics[width=0.8\linewidth]{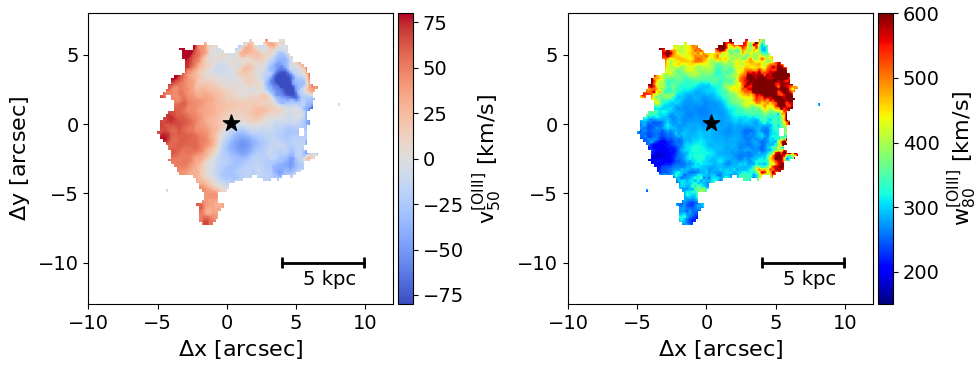}
\caption{Left panel shows the \oiii ~$v_{50}$ map and the right panel shows the \oiii ~$w_{80}$ map from two-Gaussian fits to VLT/MUSE data described in Sect. \ref{sect3.3}. The colour coding on the left panel shows \oiii ~$v_{50}$ values, ranging from -75 (blue) to 75 (red) km s$^{-1}$ and the colour coding on the right panel shows \oiii ~$w_{80}$ values, ranging from $\sim$150 (blue) to $\sim$600 km s$^{-1}$. Neither maps show ordered motions, suggesting that the \oiii ~emission traces turbulent warm ionised gas in IR20551. Outflows are apparent from high $w_{80}$ values ($>$500 km s$^{-1}$) in parts of the field. The black star marks the location of the putative AGN.}
\label{fig:oiii_w80}
\end{figure*}

\begin{figure*}
\centering
\includegraphics[width=0.8\linewidth]{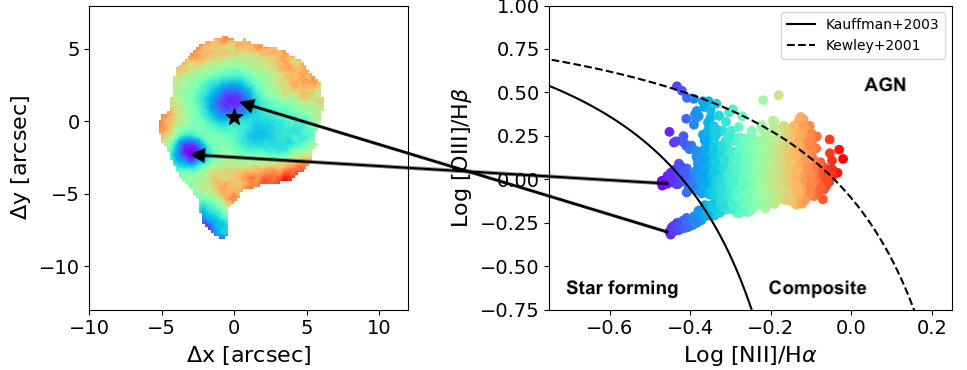}\\
\includegraphics[width=0.7\linewidth]{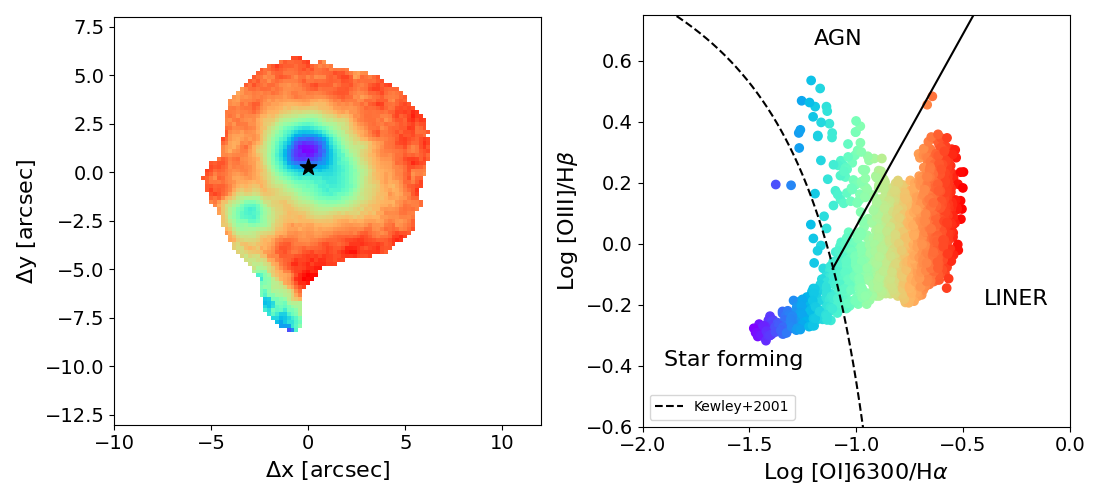}
\caption{{\it Top panels:} The left panel shows the \nii/H$\alpha$ map of IR20551 within the central $10\times 10$ arcsec$^{2}$ region and the right panel shows the corresponding \nii-BPT diagram (\oiii5007/H$\beta$ versus \nii6549/H$\alpha$). The colour coding shows the \nii6549/H$\alpha$ ratio, ranging from 0.35 (Dark Blue) to 0.9 (Red). The dashed line in the right panel indicates the line ratios for the most extreme (high-ionization) starbursts \citep{kewley01, kewley13} and the solid line shows the demarcation between the star formation and ``composite" ionisation processes (star formation+AGN) \citep{kauffmann03}. The location very close to the central nucleus (putative AGN marked by black star) is consistent with an \ion{H}{ii} region, along with another clump 5 arcseconds to the SE of the centre. The location of these two regions in the BPT diagram is shown by arrows. The bright nuclear star-forming clump accounts for the classification of IR20551 as an \ion{H}{ii} region in the literature \citep{kewley01b}. The rest of the field-of-view is dominated by composite ionisation processes. {\it Bottom panels:} The left panel shows the \oi/H$\alpha$ map of IR20551 within the central $10\times 10$ arcsec$^{2}$ region and the right panel shows the corresponding \oi-BPT diagram (\oiii5007/H$\beta$ versus \oi6300/H$\alpha$). The color coding shows the \oi6300/H$\alpha$ ratio, ranging from 0.05 (Dark Blue) to 0.25 (Red). The dashed line in the right panel indicates the line ratios for the most extreme (high-ionization) starbursts \citep{kewley01, kewley13} and the solid line shows the demarcation between the star formation and ``composite" ionisation processes (star formation+AGN).}
\label{fig:bpt}
\end{figure*}

The ALMA data cubes for CO(2-1) characterisation were already continuum-subtracted during the re-imaging procedure described in Sect. \ref{sect2.2}. To model the cold molecular gas using the CO(2-1) emission line, we follow the same procedure as modelling the H$_{2}$ emission lines in the mid-infrared spectra i.e., use multiple Gaussian functions based on the BIC criterion. Given the high spectral resolution of $\sim$4.5 km s$^{-1}$, we were able to resolve the CO(2-1) emission lines down to the asymmetries in the emission line profile. At most two Gaussians were required to reproduce the CO(2-1) line profile. The CO(2-1) spectra extracted from a circular aperture of radius 1.5 arcsec and the multi-Gaussian model is shown in Fig. \ref{fig:CO_intspec}. The radius of 1.5 arcsec was chosen to match the integrated spectrum in the MRS data. For the two-Gaussian model used here, we use a non-parametric approach which is commonly used in the literature \citep[see][ for details]{harrison14, kakkad20, wylezalek20}. We derive the following parameters with CO(2-1): $v_{50}$, $w_{80}$ ($v_{90} - v_{10}$) and the total flux of the two Gaussian profiles. $v_{\rm X}$ represents the velocity value containing X\% of flux, when integrating over the line emission from the blue-end of the spectrum. For instance, $v_{50}$ means the velocity value containing 50\% of the total flux of the emission line and is an indicator of the centroid velocity of the entire line profile. $w_{80}$ is a measure of the FWHM of the line, when using multiple Gaussian functions. The CO(2-1) line model parameters $v_{50}$ and $w_{80}$ and flux are $-90$ km s$^{-1}$ and 200 km s$^{-1}$, respectively.  We also repeat the fitting procedure for each spaxel to arrive at the maps showing the distribution of the velocity centroids, widths and the CO flux, shown in Fig. \ref{fig:CO_maps}. 

\vspace{5mm}
\noindent
We note that the spatial resolution, pixel scales and the field-of-view differ between each MRS channels, which further differ from that of the ALMA CO(2-1) observations, with which we aim to compare the warm molecular characteristics. To derive the warm H$_{2}$ excitation maps and CO-to-H$_{2}$ flux maps, it is essential that the datasets used have the same angular resolution and the pixel scales. In order to make sure that we make an accurate comparative study between two datasets, we convolve the higher resolution data with that of the lower resolution data with a Gaussian kernel and rebin the data to match the spaxel scales between the two. For example, ALMA CO(2-1) data has an angular resolution of $\sim$0.3 arcsec, while Channel-3 of the MRS, which contains the H$_{2}$ 0-0 S(2) emission line has an angular resolution of $\sim$0.8 arcsec\footnote{https://jwst-docs.stsci.edu/jwst-mid-infrared-instrument/miri-operations/miri-dithering/MIRI-MRS-psf-and-dithering}. Therefore, when deriving the flux ratio map, we convolve CO(2-1) map with a Gaussian kernel to match the angular resolution of the MRS cube, which has a lower spatial resolution.

\subsection{Resolved BPT maps with VLT/MUSE} \label{sect3.3}

The main purpose of incorporating the MUSE optical spectroscopy in this study is to determine the dominant source of ionisation (AGN, star formation, or composite) in the central region of IR20551, as inferred from emission line ratios, using a spatially resolved BPT diagrams \citep[e.g.,][]{baldwin81, veilleux87, kakkad22}, and to compare the \oiii 5007 emission-line profile, tracing diffuse warm ionised gas, with the molecular gas tracers described in Sects. \ref{sect3.1} and \ref{sect3.2}. We focus on the \nii ~and [\ion{O}{i}] ~BPT diagrams, which requires accurate measurements for the following lines: H$\beta$, \oiii$\lambda$5007, H$\alpha$, \nii$\lambda$6585 and [\ion{O}{i}]$\lambda$6300. Before carrying out the spaxel-by-spaxel fitting, we first inspect the optical spectra extracted from a 1.5 arcsec aperture (comparable to the integrated MRS and CO(2-1) spectra described in Sects. \ref{sect3.1} and \ref{sect3.2}, respectively) centered on the AGN location. This allows us to identify the detected lines and assess whether they exhibit asymmetries indicative of outflows. The integrated spectra of \oiii ~and H$\alpha$ lines are presented in the top panels of Fig. \ref{fig:muse_fluxmaps} as examples, showing tentative evidence for a blue wing in the \oiii$5007$ and H$\alpha$ line profiles.

\begin{figure*}
\centering
\includegraphics[width=0.39\linewidth]{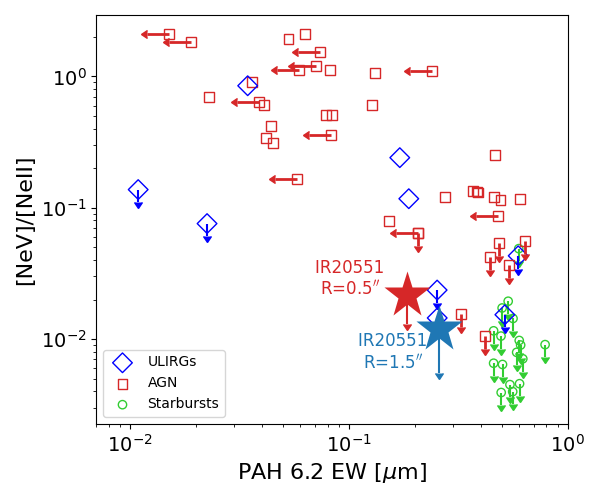}
\includegraphics[width=0.45\linewidth]{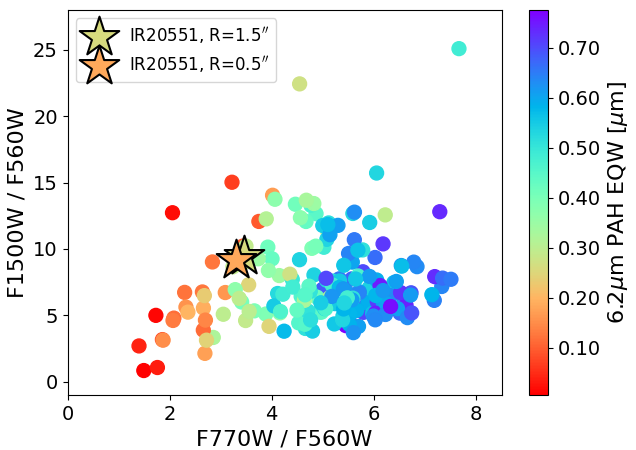}
\caption{Mid-infrared AGN diagnostics: {\it Left panel:} Flux ratio [\ion{Ne}{v}]14.3/[\ion{Ne}{ii}]12.8 versus PAH 6.2 $\mu$m equivalent width. Red squares show AGNs \citep[from][]{wu09,Tommasin2010}, blue diamonds show ULIRGs \citep{armus07} and green circles show the location of starburst galaxies \citep{Bernard-Salas09}. The location of IR20551 spectra from the two apertures, R = 0.5 arcsec and 1.5 arcsec, are shown as red and blue stars, respectively, based on upper limits on [\ion{Ne}{v}] flux. The location of IR20551 is consistent with the range expected for AGNs and ULIRGs and is somewhat shifted from the starburst regime (by a factor of 1/3). {\it Right panel:} F1500W/F560W versus F770W/F560W flux density ratio plot showing the location of the nuclear (R = 0.5 arcsec) and integrated (R = 1.5 arcsec) regions of IR20551 with respect to LIRGs in the parent GOALS sample \citep{stierwalt13}. The parent LIRG sample are colour-coded by the 6.2 $\mu$m PAH equivalent width. Its strength tends to be anti-correlated with the strength of an AGN continuum. The red and orange regions in the plot are systems dominated by AGN emission. Based on the two diagnostic diagrams shown here, we cannot rule out the presence of AGN in IR20551.}
\label{fig:MIR_colour_diagnostic}
\end{figure*}

To model these spectra, we first subtract the contribution from the stellar continuum in the optical spectra using penalised pixel fitting routine \citep[\texttt{ppxf}, see][]{cappellari04, cappellari17}, implemented within the \texttt{LZIFU} software \citep[see][]{ho16, kreckel18}. During the \texttt{ppxf} fitting procedure, we masked the emission lines, regions contaminated by sky lines, and telluric absorption features. After subtracting the stellar continuum emission, we fit the emission lines using Gaussian functions along with a linear function that takes into account any residual underlying continuum emission in the spectra.  Similar to the procedure adopted for the mid-infrared H$_{2}$ and sub-mm CO lines, the number of Gaussian functions used was decided based on the BIC values. A maximum of two Gaussians was required to reproduce the emission lines. We employ additional constraints while performing these Gaussian fits, such as setting the emission line ratio \oiii$\lambda$5007:\oiii$\lambda$4959 = 3:1 and coupling the centroid locations of all emission lines based on their expected locations in the spectra. Further details about these constraints are available in \citet{kakkad22}. 

We repeat this procedure for each spaxel to construct flux maps of H$\beta$, \oiii 5007, H$\alpha$, and \nii 6585, shown in the bottom panels of Fig. \ref{fig:muse_fluxmaps}. We also calculate $v_{50}$ and $w_{80}$ values for the \oiii ~emission line in each pixel. The \oiii ~$v_{50}$ and $w_{80}$ maps are shown in the left and right panels of Fig. \ref{fig:oiii_w80}, respectively. Blue colours in the \oiii ~$v_{50}$ map represent the blue-shifted velocities, while red colours show the red-shifted velocities. Similarly, blue colours in the \oiii ~$w_{80}$ map show regions with low velocity dispersion (close to the limit of the MUSE spectral resolution), while red colours show regions with high velocity dispersions. Since our focus is on constraining the dominant ionisation mechanism, we consider only the total line flux, without separating the narrow and broad Gaussian components. The \nii/H$\alpha$ flux ratio map is presented in the top left panel of Fig. \ref{fig:bpt}, while the distribution of individual pixels in the \nii-BPT diagram is shown in the top right panel. We also show the [\ion{O}{i}]/H$\alpha$ map and the distribution of individual pixels in the [\ion{O}{i}]-BPT diagram in the bottom panels of Fig. \ref{fig:bpt}.  We discuss these flux and velocity maps further in Sect. \ref{sect4}.

\section{Results} \label{sect4}

In this section, we present the results from the multi-wavelength imaging and spectroscopic data of IR20551, focusing on the source of ionisation, the mass and temperature of the warm molecular gas, and its comparison with the cold molecular gas mass.

\subsection{Ionisation Source in the Central Region of IR20551} \label{sect4.1}

Feedback from starbursts and/or AGN is thought to either expel a galaxy’s molecular gas or heat it, thereby suppressing future star formation. ULIRGs such as IR20551 are expected to produce strong feedback, given their high infrared luminosities, ongoing or previous mergers, and the possible presence of an AGN. In the case of IR20551, we analyse its dominant ionisation source in the central $5\times5$ arcsec$^{2}$, the area covered by the MRS. 

Rest-frame optical HST/ACS F814W and near-infrared JWST/NIRCam F277W imaging reveal the large-scale structure of IR20551, a late-stage merger with a prominent tidal tail extending SE of the nucleus and a faint tail to the NW (see top panels of Fig. \ref{fig:HST_JWST_imgs}). The NW tidal tail is better visible in the MIRI image (which is rather obscure in the HST and NIRCam images), suggesting an abundance of dust in the tidal tail. Zooming into the central $5\times5$ arcsec$^{2}$ region (bottom panels in Fig. \ref{fig:HST_JWST_imgs}), the disturbed morphology in the HST image shows features consistent with an ongoing merger. The central bright source, however, starts becoming visible in the NIRCam image, while in the MIRI image, emission from IR20551 saturates the detector. There are no indications of a second nucleus in the vicinity of the bright source based on the HST and JWST images. The extremely bright point source in the centre may come predominantly from the hot dust continuum of a buried AGN \citep{Inami2022}. 

Figure \ref{fig:nuclear_integrated_spec} shows the nuclear and integrated mid-infrared spectrum of IR20551. At shorter wavelengths, absorption features are visible around 6 $\mu$m and 6.85 $\mu$m, which may be attributed to mixture of ices and hydrogenated amorphous carbons \citep[HAC;][]{furton99, marshall07}. A deep silicate absorption feature is also visible at 9.7 $\mu$m and around 18 $\mu$m, which suggests that the central source is embedded in a smooth distribution of dust. We calculate the silicate depth at 9.7 $\mu$m, $s_{\rm 9.7~\mu m} = $ ln($f_{\lambda}/C_{\lambda}$), where $f_{\lambda}$ is the flux at the central wavelength of the 9.7 $\mu$m absorption feature and $C_{\lambda}$ is the assumed continuum level at the central wavelength in the absence of the absorption feature \citep[see][]{spoon07}. Lower values of the silicate depth imply larger absorption columns along the line of sight. We find $s_{\rm 9.7~\mu m} = $ -3.22 and -2.91 in the nuclear and integrated spectra, respectively, i.e., the absorbing column towards the central regions is larger than towards the outer regions of IR20551. The $s_{\rm 9.7~\mu m}$ values calculated here for IR20551 is also within the range observed for GOALS galaxies in \citet{stierwalt14}. These findings are also consistent with the highly obscured X-ray column density, $N_{\rm H} = 8 \times 10^{23}$ cm$^{-2}$, reported for IR20551 in the literature \citep[e.g.,][]{franceschini03}.

We detect fine structure lines in MRS spectra from ionised atoms in IR20551, such as [\ion{Ne}{ii}] at 12.8 $\mu$m, [\ion{Ne}{iii}] at 15.5 $\mu$m and [\ion{S}{iii}] at 18.7 $\mu$m. However, coronal lines such as [\ion{Ne}{v}] 14.3 $\mu$m and [\ion{O}{iv}] 25.9 $\mu$m,  produced by the hard radiation from an AGN are not detected or are very weak (see Fig. \ref{fig:Ne_specs}). The absence of these lines does not rule out the presence of an AGN, as their non-detection may result from heavy dust obscuration around the nucleus \citep[see also][]{rieke25}. Indeed, a large dust covering fraction around an AGN can also prevent the formation of Narrow Line Region and consequently, the production of emission from highly ionized atoms. We calculate an upper limit on the [\ion{Ne}{v}] 14.3 $\mu$m flux in both apertures (nuclear and integrated) and plot the flux ratio [\ion{Ne}{v}]14.3/[\ion{Ne}{ii}]12.8 versus the equivalent width of PAH 6.2 $\mu$m feature. The location of IR20551 compared to other ULIRGs, AGN and starburst galaxies is shown in the left panel in Fig. \ref{fig:MIR_colour_diagnostic}. IR20551 is located within the range expected for AGNs and ULIRGs and nearly a factor of 1/3 less than pure starbursts, which show strong PAH emission with high EWs and weak or no [\ion{Ne}{v}] emission \citep{armus07}. Even if a starburst is present, the strength of PAH 6.2 $\mu$m is reduced in the presence of hot dust. 

Mid-infrared colour-based diagnostic diagrams, such as F1500W/F760W versus F770W/F560W, can also be used to distinguish whether the central source is starburst- or AGN-dominated \citep[e.g.,][]{armus07, evans22}. In these diagrams, lower F770W/F560W ratios correspond to smaller PAH equivalent widths (EWs), indicative of AGN-dominated emission. As the MIRI image in Fig. \ref{fig:HST_JWST_imgs} is saturated, we cannot use it to calculate these colours directly. Instead, we use the available MRS spectra to construct synthetic F560W, F770W, and F1500W colours by convolving the spectra with the relevant MIRI filter throughputs. The resulting diagnostic diagram is shown in the right panel of Fig. \ref{fig:MIR_colour_diagnostic}. Targets with PAH EWs below 0.25 (shown in red/orange) are likely dominated by AGN activity \citep{armus07}. The nuclear and integrated synthetic photometric colours of IR20551 lie approximately between the regions characteristic of pure AGN and pure starburst systems, again suggesting that the contribution from an AGN cannot be excluded and that IR20551 is consistent with being a composite system.

Optical BPT diagnostics, on the other hand, tell a different story (see Fig. \ref{fig:bpt}). The central region close to the AGN has low \nii/H$\alpha$ and low \oiii/H$\beta$ emission, placing it in the star-forming region of the BPT diagram. There is a second clump towards the SE of the nucleus, also in the star-forming region of the BPT diagram, which coincides with continuum peaks in ACS/F814W and NIRCam/277W (see Fig. \ref{fig:HST_NIRCam_MIRI_Halpha}).
The lack of optical AGN emission from the galactic nucleus could result from heavy obscuration. This is consistent with the zoomed-in HST/ACS F814W in the bottom left panel of Fig. \ref{fig:HST_JWST_imgs}, which does not show a prominent central point source.

Apart from these regions, the rest of the field-of-view in IR20551 shows ``composite" emission. This indicates that not all of the forbidden line emission off the nucleus can be produced solely by \ion{H}{ii} regions. Instead, an additional contribution appears to be present.  The spiral galaxy NGC 5806 also shows a similar pattern of star-formation in its center, surrounded by extended gas with a composite AGN+ H II spectrum \citep[e.g.,][]{Robbins2025}.  This non-stellar emission can be attributed to an AGN, from either a Seyfert nucleus or a LINER \citep[see e.g.,][]{Malkan2026}.  
The most sensitive discriminator between these two kinds of non-stellar emission is the strength of the [\ion{O}{i}]$\lambda$6300 line. We, therefore, measured this line in MUSE data cube and show the spatially resolved \ion{O}{i}-BPT diagram in the bottom panels in Fig. \ref{fig:bpt}. The solid line shows the separation between Seyfert (upper) and LINER (lower) spectra. The strong [\ion{O}{i}] line emission shows that most of the observed center of the galaxy has a LINER spectrum--not that of a Seyfert AGN. Indeed spatially extended LINER emission is often observed in some regions of LIRGs.

The physical mechanism giving rise to LINER emission spectra is under debate, and could be either a very weak AGN, or ISM shocks in a wind. The LINER emission we see in IRAS20551 is not associated with the nucleus, and is, instead, found in outer regions which tend to have wider \oiii ~wings (particularly along the northern edge). We take this as weak evidence for the shock origin of the LINER emission \citep[see also][]{Sugai2000}.
 
In summary, our findings here are consistent with literature classifications of this source as an \ion{H}{ii} region at optical wavelengths. However, the AGN nature of the central source cannot be excluded with diagnostic diagrams based on mid-infrared photometry and spectra. In order to investigate whether the AGN and/or starburst drives any feedback, we perform a kinematic analysis of the emission lines tracing ionised and molecular gas in the following sections.  

\subsection{Warm molecular gas morphology and kinematics} \label{sect4.2}

We detect the following H$_{2}$ rotational transitions in the MRS spectrum of IR20551: 0-0 S(1), 0-0 S(2), 0-0 S(3), 0-0 S(4), 0-0 S(5), 0-0 S(6), 0-0 S(7) and 0-0 S(8) observed at wavelengths 17.767, 12.807, 10.081, 8.370, 7.207, 6.371, 5.748 and 5.270 $\mu$m, respectively. We first summarise the H$_{2}$ detections in the nuclear and integrated spectra, before diving into the spatially-resolved properties from the spaxel-by-spaxel analysis. The nuclear and integrated H$_{2}$ spectra, along with the respective Gaussian models are shown in Fig. \ref{fig:h2_specs}. Figure \ref{fig:h2_maps} shows the line centroid (left panels), width (FWHM, middle panel) and flux maps (right panel) of S(2) and S(7) transitions (top and bottom panels, respectively) as examples.

The extinction corrected fluxes of the individual H$_{2}$ lines are reported in Table \ref{tab:h2_lines} and are in the range $1.1-10.3 \times 10^{-14}$ erg s$^{-1}$ cm$^{-2}$ in the nuclear spectra and $1.6-26.9 \times 10^{-14}$ erg s$^{-1}$ cm$^{-2}$ in the integrated spectra. The extinction correction factors were obtained from \texttt{CAFE}. The line flux maps shown in the right panels of Fig. \ref{fig:h2_maps} clearly show two prominent tidal tails in this system due to the ongoing merger: one towards the SE of the nucleus and another towards the NE. The NE tidal tail is not visible in the maps of higher transitions (S(3)--S(8)) due to the smaller field-of-view in the lower channels of MRS.

\begin{figure}
\centering
\includegraphics[width=0.8\linewidth]{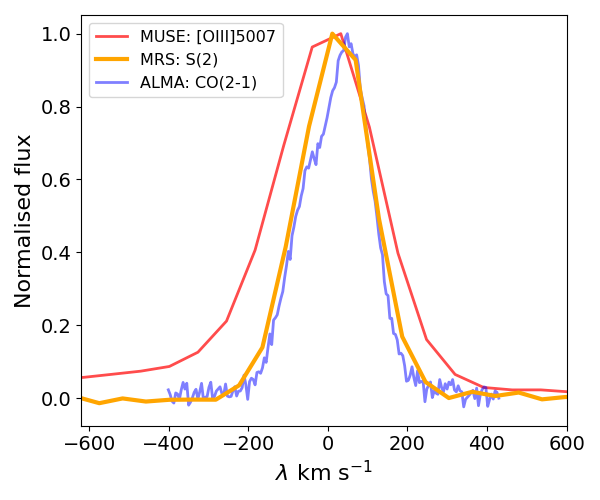}
\caption{Normalised \oiii$\lambda$5007, H$_{2}$ S(2), and CO(2–1) line profiles plotted on the same wavelength scale. The \oiii ~line exhibits the largest velocity dispersion with $w_{80}^{\rm [OIII]} \sim 790$ km s$^{-1}$. While the CO(2–1) profile shows mild asymmetry (see also Fig. \ref{fig:CO_intspec}), its overall width is comparable to that of the S(2) transition. These profiles point to the presence of warm ionised outflows in IR20551, whereas the cold and warm molecular gas phases show only slow-moving gas or no evidence of outflows.}
\label{fig:OIII_CO_S2_spec}
\end{figure}

The H$_{2}$ line centroid positions (based on the redshifts derived from optical lines -- \oiii and H$\alpha$ -- in MUSE spectra) for each transition (see Table \ref{tab:h2_lines}) show no significant variation between the two apertures, with velocity differences remaining below $\Delta v < 15$ km s$^{-1}$. A comparison of the H$_{2}$ line flux maps with the centroid maps suggests that the NE tidal tail may be tracing a gas inflow toward the AGN, as indicated by the redshifted velocities seen in the left panels of Fig. \ref{fig:h2_maps}. The centroid maps further reveal a nearly continuous velocity gradient extending from the NE to the SW of the nucleus, with the strongest redshifts located in the NE tidal tail. This relatively smooth gradient is likely a consequence of IR20551 being in a late merger stage, where the circumnuclear region begins to settle into a rotating disk.

\begin{figure*}
\centering
\includegraphics[width=0.7\textwidth]{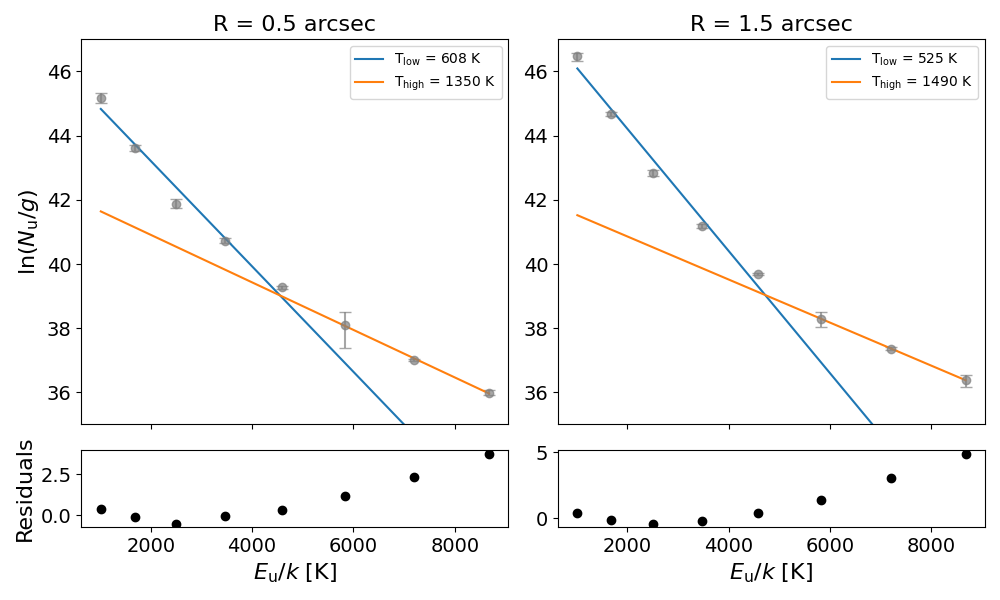}\\
\includegraphics[width=0.7\linewidth]{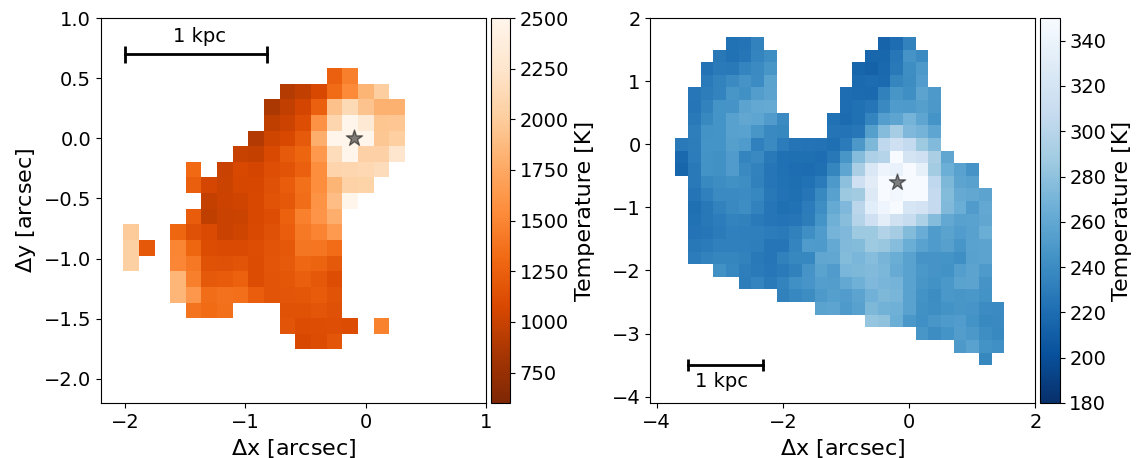}
\caption{{\it Top panels:} Temperatures derived from the nuclear (top left) and integrated (top right) spectra. The nuclear spectrum is extracted from a 0.5 arcsec circular aperture, while the integrated spectrum is taken from a 1.5 arcsec aperture, both centred on the AGN. The plots illustrate a two-temperature model for the H$_{2}$ transitions: a warm component at $\approx$500 K, modelled using the S(1)–S(3) lines, and a hot component at $\approx$1400 K, traced by the S(4)–S(8) lines. {\it Bottom panels:} The maps show the spatial distributions of the molecular gas temperature based on the two-temperature excitation models shown in the top panels. The bottom left panel shows the higher temperature and the bottom right panel shows the lower temperature distribution. The gas temperature is highest at the location of the putative AGN, consistent with it being heated by accretion activity.}
\label{fig:Tvib_plots}
\end{figure*}

All the H$_{2}$ transitions are spectrally resolved, with LSF-corrected line widths (FWHM) in the range $\sim$180–230 km s$^{-1}$. The median widths in the nuclear and integrated spectra are comparable, at 192 km s$^{-1}$ and 190 km s$^{-1}$, respectively. The H$_{2}$ profiles in the MRS spectra are well described by a single Gaussian component and show no evidence of extended blue or red wings in either aperture (Fig. \ref{fig:h2_specs}). This contrasts with the \oiii ~and CO(2–1) profiles, which exhibit asymmetric shapes. Figure \ref{fig:OIII_CO_S2_spec} compares the \oiii, CO(2–1), and S(2) lines, each normalised and plotted on the same velocity scale relative to its centroid. Among these, \oiii ~displays the broadest profile, with $w_{80}^{\rm [OIII]} > 500$ km s$^{-1}$, suggesting the presence of outflows in the warm ionised gas phase. The presence of outflows or turbulent motion in the ionised gas is further evident from the maps in the right panel of Fig. \ref{fig:oiii_w80}, where the \oiii ~$w_{80}$ values reach $>$500 km s$^{-1}$, values that cannot be explained by simple rotation. Moreover, the absence of clear structure or a coherent velocity gradient across the $w_{80}$ map also suggests turbulence in the ionised gas. We also calculate mass outflow rates in the centre of IR20551 (using the MUSE aperture spectrum shown in Fig. \ref{fig:muse_fluxmaps}) following the method described in \citet{kakkad22}, assuming solar metallicity of the gas and calculating the electron density from the \sii$\lambda\lambda$6716, 6731 doublet. Assuming that the broad Gaussian component represents the outflowing gas, we find an ionised gas mass outflow rate of $<$0.01 M$_{\odot}$ yr$^{-1}$. The CO(2–1) line (blue curve in Fig. \ref{fig:OIII_CO_S2_spec}) shows mild asymmetry and requires two Gaussian components (Fig. \ref{fig:CO_intspec}), though the individual Gaussian components have modest widths of $\sim$120 km s$^{-1}$ and $\sim$140 km s$^{-1}$. These asymmetries may simply reflect turbulence in the central regions or slower moving gas compared to the ionised phase. Visually, the S(2) line in the MRS spectra has a width comparable to CO(2–1) but lacks its asymmetry, likely due to the superior velocity resolution and spectral sampling of the ALMA data. A broader discussion of these line widths in the context of outflows is presented in Section \ref{sect5}.

%{\bf TBC: Comparison with PAH maps? - See Fig. \ref{fig:PAH}, repeated from the caption here: PAH emission does not seem to show the same structure as the rotational transitions: e.g., the tail is absent. PAH emission appears more symmetric. If this is contribution from starburst, it validates the MUSE BPT map, which shows contribution from star forming regions in the centre. Note that this is an approximate map using a single Gaussian fit.}

\subsection{H$_{2}$ excitation in IR20551} \label{sect4.3}

The multiple H$_{2}$ transitions detected in the MRS spectra allow us to construct a rotation diagram of the warm molecular hydrogen, from which we can derive both the excitation temperature and, subsequently, the mass of the warm molecular gas. Warm molecular gas typically exhibits excitation temperatures of a few hundred K, making these mid-infrared transitions particularly well suited to probing CO-dark molecular gas that are not detectable at sub-mm wavelengths. In this section, we examine the temperature of the warm gas in the nuclear and integrated spectra, estimate the total warm molecular gas mass, and use the spatially resolved data to investigate the temperature distribution across the system.

We created the rotational diagrams from the line fluxes in the nuclear and integrated spectra, following the methodology in \citet{rigopoulou02} \citep[see also][]{zakamska10, togi16, petric18}. The top panels in Figure \ref{fig:Tvib_plots} shows the molecular hydrogen excitation diagram of warm hydrogen, based on the H$_{2}$ line detections in the nuclear spectrum (top left panel) and the integrated spectrum (top right panel). In making these rotational diagrams, we assume the gas to be in Local Thermodynamic Equilibrium (LTE) and an ortho-para ratio of 3:1. Compared to the mid-infrared spectra of IR20551 published so far in the literature \citep[e.g.,][]{higdon06, sani12}, the H$_{2}$ transitions reported in this paper probe a larger range in the upper energy levels ($E_{\rm u}/k \sim 1000-9000$ K). The excitation temperature is the inverse of the slope of the excitation diagram. 

It is clear from the excitation diagram in Fig. \ref{fig:Tvib_plots} that the warm molecular gas is not well represented by a single temperature, but a range of temperatures. This is not surprising given that the molecular gas generally consists of various components at different temperatures, especially in ULIRGs which are complex systems consisting of ionisation by AGN, star formation/starbursts and shocks. The nuclear and integrated MRS spectra of IR20551 broadly fall into two temperature distributions, shown by the blue and orange lines in the top panels in Fig. \ref{fig:Tvib_plots}: One a warm molecular gas component at an excitation temperature of $\sim$608 K and $\sim$525 K  and a hot gas component at an excitation temperature of $\sim$1350 K and $\sim$1490 K in the nuclear and integrated spectra, respectively. The hotter phase of the gas likely probe similar gas phase as the ones traced by the ro-vibrational H$_{2}$ transitions in the near-infrared wavelengths \citep[e.g.,][]{rosario19, riffel23, bianchin24}.

In contrast to a two temperature model, we also used a more physical approach that models a continuous gas temperature distribution to determine warm gas mass. We fit the rotational H$_{2}$ line fluxes with a power-law, temperature dependent excitation function \citep[see][for details]{togi16}. The \citet{togi16} model assumes an upper bound on the temperature distribution at 2000 K, as the mass contribution from rotational transitions beyond this temperature is negligible. We also assume an ortho-para ratio of 3 for the statistical weights of the molecular transitions. This ratio is known to drop significantly at lower temperatures \citep[T$\leq$30K, see][]{burton92}. However, for the warm gas temperatures considered here (T$\geq$200 K), this assumption provides a useful conservative upper limit. We find a power-law index, $n = 4.76\pm0.16$, consistent with the range observed in ULIRGs \citep[see][]{togi16} and approximately $6.1\times10^{8}$ M$_{\odot}$ of warm molecular gas is observed at T$\geq$200 K in aperture R = 1.5 arcsec. We will compare this warm molecular gas mass with the cold molecular gas mass in Sect. \ref{sect4.4}.

We take advantage of the spatially-resolved IFU data from the MRS to derive excitation maps of the warm and the hot molecular gas components. The H$_{2}$ excitation maps of the warm and the hot gas are shown in the bottom panels in Fig. \ref{fig:Tvib_plots}. These maps clearly show a peak in the temperature distribution corresponding to the location of the putative AGN, with a slightly higher temperature in the NE tidal tail. The temperature distribution likely suggests that the putative AGN radiation may be the source of ionisation of the warm molecular gas, however the extended warm gas may be excited by non-AGN processes such as shocks from ionised outflows (see Sect. \ref{sect5}).

\subsection{Cold gas to warm gas ratios} \label{sect4.4}

\begin{figure}
\centering
\includegraphics[width=0.8\linewidth]{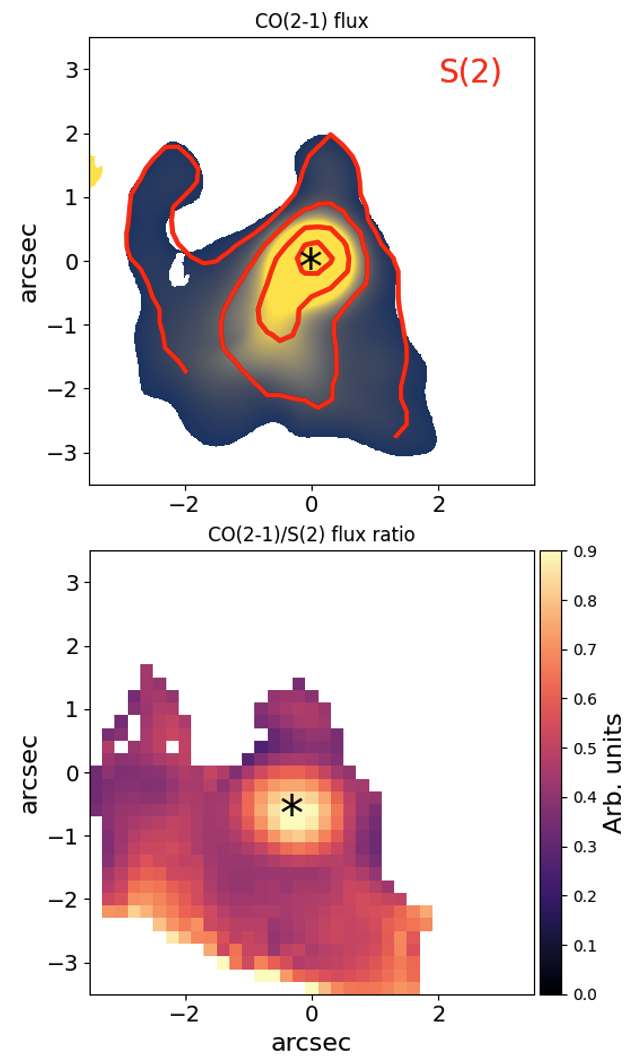}
\caption{{\it Top panel:} The map in this plot shows the CO(2-1) flux distribution, same as the right panel in Fig. \ref{fig:CO_maps}. Emission from the S(2) line tracing the warm molecular gas is overlaid as red contours at levels 10\%, 20\%, 50\% and 90\% of the peak flux value. The map shows that the cold and warm gas are co-located in the central regions of IR20551. {\it Bottom panel:} Ratio between the fluxes of CO(2-1), tracing the cold molecular gas and S(2) transition, tracing the warm molecular gas, on an arbitrary scale. The CO-based cold molecular gas dominates the mass across the field-of-view. The maximum ratio of cold-to-warm gas is observed in the centre, consistent with an ongoing star formation in this region.}
\label{fig:co_h2_overlay}
\end{figure}

Most molecular gas studies in the literature focus on the cold component (T$<$100 K), typically traced via CO transitions at sub-mm wavelengths. The combination of ALMA and JWST data offers a unique opportunity to examine molecular gas abundance ratios across different temperatures, both globally and in a spatially resolved manner near the AGN. In this section, we first characterise the CO(2–1)-based cold molecular gas (total mass, morphology, and kinematics) and compare it with the warm molecular gas traced by the S(2) transition. We chose the S(2) line for comparison as its location in channel-3 of MRS provides a larger field of view than the higher-excitation warm gas transitions in the lower MRS channels.

We first compare the integrated (aperture radius = 1.5 arcsec) cold and warm molecular gas masses. We derived a warm molecular gas mass of $\sim 6.1 \times 10^{8}$ M$_{\odot}$ in Sect. \ref{sect4.3}. To calculate the cold molecular gas mass from CO(2-1) line, we use the following equation from \citet{solomon05}: 

\begin{equation}
L^{\prime}_{\rm CO} = 3.25\times10^{7} ~S_{\rm CO} ~\Delta v ~\nu_{\rm obs}^{-2} ~D_{\rm L}^{2} ~(1+z)^{-3}
\end{equation}

where $L^{\prime}_{\rm CO}$ is the luminosity of the CO transition measured in K km s$^{-1}$ pc$^{2}$, $S_{\rm CO} ~\Delta v$ is the velocity-integrated flux in Jy km s$^{-1}$, $\nu_{\rm obs}$ is the observed frequency of the CO(2-1) line ($\sim$221 GHz) and $D_{\rm L}$ is the luminosity distance. From the integrated CO transition shown in Fig. \ref{fig:CO_intspec}, we find $L^{\prime}_{\rm CO(2-1)} = 2.8\times10^{9}$ K km s$^{-1}$ pc$^2$. Assuming an excitation correction, $r_{21} = 0.4$ \citep[see][]{iono09, papadopoulos12, bolatto13}, we get a cold molecular gas mass of $M_{\rm H_{2}}^{\rm cold} = 1.3 \times 10^{10}$ M$_{\odot}$, using an $\alpha_{\rm CO}$ conversion factor of  1.8 M$_{\odot}$ K$^{-1}$ km$^{-1}$ s pc$^{-2}$, which is consistent with the values found for GOALS galaxies in \citet{herrero-illana19} and within the range for ULIRGs in \citet{bolatto13}. The ratio of cold-to-warm molecular gas mass is, therefore, approximately 20. In other words, $>$95\% of the molecular gas is in cold gas phase. This suggests that the central regions of IR20551 still has a vast reservoir of cold gas to continue forming stars. However, there is a non-negligible fraction of warm molecular gas ($<$5\%), possibly excited by the putative AGN and shocks (based on the excitation temperature map in the bottom panels in Fig. \ref{fig:Tvib_plots}. 

% https://ui.adsabs.harvard.edu/abs/2019A%26A...628A..71H/abstract --> alpha_CO = 1.8 for GOALS galaxies. 

We use the spatially resolved data from ALMA and MRS to compare the morphology and kinematics of the cold and warm molecular gas, respectively. Figure \ref{fig:CO_maps} presents the centroid velocity (left), line width (FWHM, middle), and flux (right) maps of the CO(2–1) line from the ALMA observations of IR20551. The cold gas distribution and kinematics traced by CO(2-1) are similar to that of the warm gas traced by the H$_{2}$ rotational transitions in the MRS data (Fig. \ref{fig:h2_maps}). For instance, the CO(2-1) flux map also reveals a prominent tidal tail extending SE of the AGN, and the centroid velocity field shows a similar structure and orientation to that of the warm molecular gas. A direct comparison of the CO(2–1) and S(2) distributions is shown in Fig. \ref{fig:co_h2_overlay}, where the background shows the CO(2–1) emission with S(2) emission overlaid in red contours. The figure clearly demonstrates that the cold and warm molecular gas are co-spatial. Unlike some resolved multi-phase studies reported in the literature, we do not find regions in IR20551 devoid of cold molecular gas \citep[e.g.,][]{rosario19, davies24}.

The bottom panel in Fig. \ref{fig:co_h2_overlay} shows the ratio between the flux of the CO(2-1) and S(2) line (normalised to an arbitrary scale). Note that while deriving this flux ratio map, we ensured that the spatial resolution and sampling of the two datasets are consistent with each other, as described in Sect. \ref{sect3.2}. We do not convert the CO(2-1) and S(2) fluxes to molecular gas masses, as the purpose of these plots is to estimate the spatial variation in the cold-to-warm gas ratio rather than deriving the exact ratios, which are affected by several assumptions such as conversion ratios or excitation corrections (which also show spatial variation, see discussion in Sect. \ref{sect5}). There is a higher fraction of cold molecular gas compared to warm gas in the centre of IR20551 and the flux ratio drops off as we move away from the centre by a factor of $\sim$3 as we move towards the NE tidal tail. Despite the drop, the cold gas dominates the flux (mass) budget across the targeted field-of-view, which might be fuelling the central starburst (visible in the optical wavelengths) as well as the putative AGN. The spatially-resolved dominance of cold molecular gas is also consistent with the integrated spectral measurements discussed earlier in this section.

\section{Discussion} \label{sect5}

ULIRGs such as IR20551 provide ideal laboratories for studying the effects of feedback from starbursts and/or AGN on the host galaxy environment. The classification of IR20551 as AGN or starburst-dominated system depended on the wavelength band of observation. X-ray observations yield mixed results: Chandra detected the source \citep{ptak03}, XMM-Newton identified the source as hosting an obscured AGN \citep{franceschini03}, while NuSTAR detected only a marginal (2$\sigma$) signal in the 8–24 keV band \citep[see][]{yamada21}. Despite the weak NuSTAR detection, the high column density reported in this system supports the presence of an obscured AGN \citep{franceschini03}. In contrast, optical spectroscopy previously classified IR20551 as an \ion{H}{ii} region (star forming) galaxy \citep[e.g.,][]{kewley01b}. Our spatially resolved MUSE data enable the construction of a BPT map, which confirms \ion{H}{ii}-like ionisation close to the central region of IR20551 in optical wavelengths (Fig. \ref{fig:bpt}). The galaxy hosts two star-forming clumps, visible as the blue regions in the left panel of Fig. \ref{fig:bpt}. However, the remainder of the field-of-view exhibits composite ionisation, and in some off-nuclear regions even AGN-like signatures -- although the line ratios observed in the \nii ~and [\ion{O}{i}] BPT diagrams show that the emission could also be associated with LINER emission from ISM shocks in a wind due to IR20551 being a late-stage merger.

The AGN nature of the nucleus cannot be ruled out based on diagnostic plots at mid-infrared wavelengths. While this source was previously classified as a starburst in the mid-infrared \citep[e.g.,][]{genzel98}, the mid-infrared spectral diagnostics ([\ion{Ne}{v}]/[\ion{Ne}{ii}] versus PAH 6.2 $\mu$m EW) and colour–colour diagnostics (F1500W/F560W versus F770W/F560W) based on synthetic photometry extracted from the MRS spectra, places IR20551 at the boundary of the AGN-dominated and star forming regimes (see Fig. \ref{fig:MIR_colour_diagnostic}). The MIRI imaging could not be used directly for this diagnostic as it is saturated in the centre. High-ionisation emission lines such as [\ion{Ne}{v}], which are characteristic tracers of AGN activity, are either weak or absent in the MRS spectra. Their absence may instead reflect the heavy obscuration in this system, as indicated also by the prominent 9.7 $\mu$m silicate absorption feature. Overall, based on mid-infrared diagnostics, the observations do not rule out the presence of an obscured AGN in IR20551, along with at least two central star-forming clumps, showing that the galaxy is also actively forming stars.

We investigated whether the AGN in IR20551 has a potential for any feedback on its host galaxy through outflows. Such outflows can, in principle, deplete the gas reservoir available for star formation and eventually quench star formation over long timescales. To test this, we examined the emission-line profiles of the ionised, warm, and cold molecular gas using MUSE, MRS, and ALMA spectra, respectively, searching for asymmetric or broad features. The integrated (aperture radius = 1.5 arcsec) normalised spectra of \oiii, S(2), and CO(2–1), plotted on the same velocity scale in Fig. \ref{fig:OIII_CO_S2_spec}, reveal that the warm ionised gas traced by \oiii ~shows the broadest profile, with $w_{80} \sim 790$ km s$^{-1}$. The turbulent nature of the ionised gas is also visible in the $v_{50}$ and $w_{80}$ maps of \oiii ~presented in Fig. \ref{fig:oiii_w80}, which shows non-ordered motions. In contrast, the integrated CO(2–1) profile, although asymmetric and requiring two Gaussian components to reproduce, has a significantly narrower width of $w_{80} \sim 210$ km s$^{-1}$. The CO $v_{50}$ maps at least shows a relatively smooth gradient in the spatial profile compared to \oiii. This may suggest that the cold molecular gas moves more slowly than the ionised phase, possibly tracing the disk which is in the process of settling, and the observed asymmetries may simply reflect turbulent motions in the central regions, while the ionised gas is tracing the Narrow Line region. The S(2) warm H$_{2}$ transition shows a width comparable to CO(2–1) but no evidence of asymmetric wings. While the absence of wings in the warm H$_{2}$ line could be partly due to the coarser spectral sampling of the MRS data, the current observations clearly suggest that neither molecular gas phase exhibits high-velocity outflows.

The absence of molecular gas outflows in IR20551 contrasts with numerous studies of AGN host galaxies, ULIRGs, and starbursts using Spitzer/IRS and JWST/MRS, where warm molecular outflows with velocity dispersions exceeding 400 km s$^{-1}$ have been reported \citep[e.g.,][]{dasyra11, bohn24, costa-souza24, dan25, zanchettin25}. However, the detection of outflows in a given gas phase strongly depends on factors such as the timescale of AGN-driven feedback, the efficiency of its coupling to the ISM, and the prevailing ISM conditions \citep[see also][]{harrison24}. Physically, \oiii-based ionised gas originates mainly in the extended NLR, outside the galactic disk, making high-velocity outflows more readily detectable in this phase. By contrast, molecular gas resides predominantly in the disk, where any outflows are expected to be slower or their wings obscured by dust. Indeed, the H$_{2}$ line-width maps reveal turbulence within the central $5\times5$ arcsec$^{2}$ region, but the resolved widths remain around $\sim$200 km s$^{-1}$, which do not suggest a fast outflow in the molecular gas phases. 

The excitation mechanism of the warm H$_{2}$ transitions is challenging to pin down, as they may arise from a variety of processes including shocks, radio jets, or external UV radiation. The excitation temperature maps in Fig. \ref{fig:Tvib_plots} show a peak near the putative AGN location, suggesting that UV radiation from the accretion disk likely dominates the heating of warm H$_{2}$ in the central region. A similar conclusion was reached for the nearby LIRG NGC 3256, based on H$_{2}$ line ratios \citep{bohn24}. As mentioned earlier, IR20551 also exhibits ionised gas outflows, pointing to the possible contribution of shocks in exciting H$_{2}$. Indeed, the models of \citet{kristensen23} suggest that the observed H$_{2}$ fluxes in IR20551, at least in the regions away from the AGN, can be explained with shock excitation. Radio jets have also been proposed as a mechanism for H$_{2}$ excitation \citep[e.g.,][]{davies24}, although the available radio data are at much lower spatial resolution than the MRS and ALMA data \citep[see][]{condon21}. Based on the data presented here, the putative AGN likely plays a key role in powering the excitation of the warm H$_{2}$ transitions.

\begin{figure}
\centering
\includegraphics[width=0.8\linewidth]{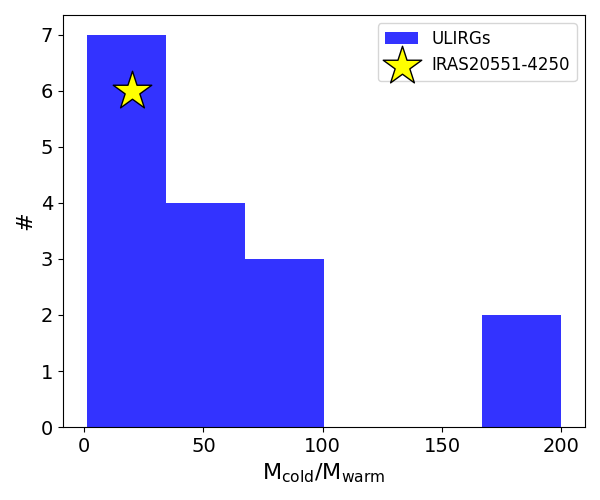}
\caption{Distribution of cold-to-warm gas ratio in low-redshift ULIRGs from the literature. Warm gas masses in this distribution were obtained from the sample in \citet{petric18} and cold molecular gas masses from \citet{alatalo24}. The observed cold molecular gas fraction of IR20551 is consistent with the range observed for low redshift ULIRGs.}
\label{fig:gas_ratio_lit_comparison}
\end{figure}

Finally, star formation in galaxies predominantly occurs in regions rich in cold molecular gas, and feedback from stars and/or AGN can act in either an ejective or preventative mode i.e., removing gas from the host galaxy or suppressing the ability of the available gas to form stars. In IR20551, we find fast ionised outflows but  no evidence of an outflow in molecular gas phase or that the molecular gas is ejected. The warm molecular gas contributes $<$5\% of the total molecular reservoir. The warm gas is likely excited by AGN radiation and therefore, has the potential to play a role in preventative feedback over longer timescales. However, the cold molecular phase still remains dominant in IR20551. Similar cold-to-warm molecular gas fractions have been reported in outflows by \citet{bohn24}. We compiled cold and warm molecular gas masses for a sample of low-redshift ULIRGs and derived their cold-to-warm molecular gas fractions, as shown in Fig. \ref{fig:gas_ratio_lit_comparison}. The warm molecular gas masses presented in Fig. \ref{fig:gas_ratio_lit_comparison} are taken from \citet{petric18}, based on a sub-sample of the parent GOALS galaxies from where IR20551 was selected. In \citet{petric18}, warm gas masses are derived from integrated measurements of high-resolution Spitzer/IRS spectra, using the same methodology adopted in this paper. The cold molecular gas masses are drawn from integrated CARMA CO(1–0) observations \citet{alatalo24}, who assume an $\alpha_{\rm CO}$ value of 1.5, comparable to the value of 1.8 used in this paper for IR20551. We find that the cold-to-warm molecular gas ratio in IR20551 is broadly consistent with those reported in the literature for low-redshift ULIRGs. However, we note that the cold gas mass estimate is subject to systematic uncertainties from assumed $\alpha_{\rm CO}$ value, which relies on the physical conditions of the ISM such as shocks, photo-dissociated regions, metallicity etc \citep[see][]{narayanan12, bolatto13}. Furthermore, these conditions change spatially depending on the local ISM conditions. Therefore, the flux ratio map between CO(2-1) and S(2) shown in Fig. \ref{fig:co_h2_overlay} is under the assumption of a uniform $\alpha_{\rm CO}$ value applied across the field-of-view.

% As we wanted to obtain a first-order look at the gas ratios in comparable galaxy, we do not attempt to make aperture corrections in making these comparisons. 

In IR20551, the abundance of cold molecular gas reservoir ($M_{\rm H_2}^{\rm cold} \sim 1.3\times10^{10}$ M$_{\odot}$) suggests that star formation can continue largely unaffected, and feedback from the central source has not yet significantly altered the star-forming cycle. Although IR20551 hosts an ionised outflow, a mass outflow rate of $<$0.01 M$_{\odot}$ yr$^{-1}$ in the centre does not appear sufficient to expel the molecular gas or suppress star formation on short timescales. We also do not see any evidence of enhancement of star formation due to outflows or turbulent motions \citep[e.g.,][]{cresci15b, bessiere22, mercedes-feliz24}. The H$\alpha$ clump towards the SE of the putative AGN (Figs. \ref{fig:muse_fluxmaps} and \ref{fig:HST_NIRCam_MIRI_Halpha}) happens to be at a location where the warm and cold molecular gas split into the tidal tails. However, this  is outside the field-of-view of MRS observations and we cannot conclude whether this star forming clump is indeed triggered due to warm or cold gas flows.

We note that additional phases, such as hot molecular gas traced by ro-vibrational transitions in the near-infrared ($T > 1000$ K), also contribute to the molecular reservoir. However, this component generally represents only a small fraction compared to the warm and cold phases \citep[e.g.,][]{kakkad25, zanchettin25}. Thus, excluding the hot molecular gas from our analysis is unlikely to significantly impact the results presented in this paper. 

\section{Summary and future outlook}

In this paper, we presented multi-wavelength, spatially resolved observations of IRAS20551-4250, a ULIRG at redshift $z=0.0429$, using imaging and spectroscopy from JWST (MIRI, MRS and NIRCam), HST (ACS), VLT (MUSE) and ALMA. Our goal was to characterise the warm and cold molecular gas within the central $5\times5$ arcsec$^{2}$ ($\sim4\times4$ kpc$^{2}$) of the galaxy and assess the impact of the AGN on the multi-phase molecular gas properties. To this end, we first examined whether AGN ionisation dominates at the observed wavelengths, followed by a detailed characterisation of the warm molecular gas traced using the rotational transitions in the mid-infrared spectra (temperature, mass, distribution, kinematics), which we then compared with the cold gas properties derived from ALMA observations. Here, we summarise the main findings of this paper: 

\begin{itemize}
\item The wide field images of IRAS20551-4250 show tidal tails. At optical wavelengths, the central region of the galaxy shows a disturbed morphology, but as we move to the near-infrared and mid-infrared wavelengths, a central bright source becomes prominent. Spectral and colour-colour diagnostics in the mid-infrared wavelengths does not rule out the AGN nature of the central source. The MRS spectra shows a deep silicate feature which suggests a possible presence of an obscured AGN. The galaxy also hosts a starburst, confirmed via BPT diagrams constructed using optical IFU observations from MUSE, confirming the composite nature of the source as reported in previous literature.  
\item We detect several rotational H$_{2}$ transitions in the MRS spectra, from S(1) to S(8). The excitation diagram of these transitions suggest that the molecular gas can be approximately grouped into two temperatures: one warm molecular gas at $\sim$500 K and other hot gas at $\sim$1400 K. The excitation map suggests that the putative AGN may be the likely source for exciting the rotational transitions, but the observed warm gas fluxes could also be explained using shocks. 
\item Fast outflows with velocity dispersion $\sim$790 km s$^{-1}$ and mass outflow rate of $<0.01$ M$_{\odot}$ yr$^{-1}$ are observed in the \oiii5007-based ionised gas phase, while the cold and warm molecular gas, with velocity dispersions $\sim$200 km s$^{-1}$ show slow moving or no outflows.
\item The warm and cold molecular gas are co-located and show similar kinematics. Both gas phases show two tidal tails and the central region of this galaxy appears to be in the process of settling into a rotating disk. 
\item We find a warm molecular gas mass of $6.1\times10^{8}$ M$_{\odot}$ and cold molecular gas of $1.3\times10^{10}$ M$_{\odot}$, i.e., warm molecular gas constitutes $<$5\% of the total molecular gas in this galaxy. The cold-to-warm molecular gas fraction is consistent with literature measurements of low redshift ULIRGs. The higher amount of cold molecular gas suggests that the galaxy still has sufficient amount of cold molecular gas to continue forming stars in the centre. The ionised gas mass outflow rate of $<0.01$ M$_{\odot}$ yr$^{-1}$ appear insufficient to immediately disrupt the star forming ability of IR20551.
\end{itemize}

IRAS20551-4250, with its observed cold-to-warm molecular gas ratios, ongoing star formation, and the presence of ionised outflows, represents a ULIRG that hosts an ionised outflow, indicated by high \oiii ~velocity dispersion values, but the outflow is not sufficient to develop the wide-scale feedback needed to impact the galaxy’s star-forming capacity. Expanding such analyses with JWST/MRS and ALMA observations of a larger sample of low-redshift ULIRGs and AGN hosts will provide the means to assess more generally whether observations of outflows and subsequent feedback from the AGN is a common occurence or do we expect to see more galaxies like IRAS20551-4250 where the AGN appears to have minimal impact on the star formation cycle. 

\section*{Acknowledgements}
We thank the anonymous referee for their constructive comments and suggestions. CR acknowledges support from SNSF Consolidator grant F01$-$13252, Fondecyt Regular grant 1230345, ANID BASAL project FB210003 and the China-Chile joint research fund. V.U acknowledges funding support from NSF Astronomy and Astrophysics Grant (AAG) No. AST-2536603, NASA Astrophysics Data Analysis Program (ADAP) grant No. 80NSSC23K0750, NASA Astrophysics Decadal Survey Precursor Science (ADSPS) grant No. 80NSSC25K0169, and STScI grant Nos. HST-GO-17285.001-A and JWST-GO-01717.001-A. MSG acknowledges support by the Hellenic Foundation for Research and Innovation (HFRI) under the '2nd Call for HFRI Research Projects to support Faculty Members \& Researchers' (Project Number: 03382).This paper makes use of the following ALMA data: ADS/JAO.ALMA\#2022.1.01262.S. ALMA is a partnership of ESO (representing its member states), NSF (USA) and NINS (Japan), together with NRC (Canada), NSC and ASIAA (Taiwan), and KASI (Republic of Korea), in cooperation with the Republic of Chile. The Joint ALMA Observatory is operated by ESO, AUI/NRAO and NAOJ. This research is based on observations made with the NASA/ESA Hubble Space Telescope obtained from the Space Telescope Science Institute, which is operated by the Association of Universities for Research in Astronomy, Inc., under NASA contract NAS 5–26555. These observations are associated with program 10592. This work is based [in part] on observations made with the NASA/ESA/CSA James Webb Space Telescope. The data were obtained from the Mikulski Archive for Space Telescopes at the Space Telescope Science Institute, which is operated by the Association of Universities for Research in Astronomy, Inc., under NASA contract NAS 5-03127 for JWST. These observations are associated with program 3368. 

%%%%%%%%%%%%%%%%%%%%%%%%%%%%%%%%%%%%%%%%%%%%%%%%%%
\section*{Data Availability}

The data can be retrieved from MAST archive under the programme IDs 10592 (HST) and 3368 (JWST). ALMA data are available from the ALMA archive under the programme ID 2022.1.01262.S and the MUSE data are available under the programme ID 095.B-0049.
%%%%%%%%%%%%%%%%%%%% REFERENCES %%%%%%%%%%%%%%%%%%

% The best way to enter references is to use BibTeX:

\bibliographystyle{mnras}
\bibliography{reference} % if your bibtex file is called example.bib

% Alternatively you could enter them by hand, like this:
% This method is tedious and prone to error if you have lots of references
%\begin{thebibliography}{99}
%\bibitem[\protect\citeauthoryear{Author}{2012}]{Author2012}
%Author A.~N., 2013, Journal of Improbable Astronomy, 1, 1
%\bibitem[\protect\citeauthoryear{Others}{2013}]{Others2013}
%Others S., 2012, Journal of Interesting Stuff, 17, 198
%\end{thebibliography}

%%%%%%%%%%%%%%%%%%%%%%%%%%%%%%%%%%%%%%%%%%%%%%%%%%

%%%%%%%%%%%%%%%%% APPENDICES %%%%%%%%%%%%%%%%%%%%%

\appendix

\section{Appendix}

\begin{figure*}
\centering
\includegraphics[width=0.9\linewidth]{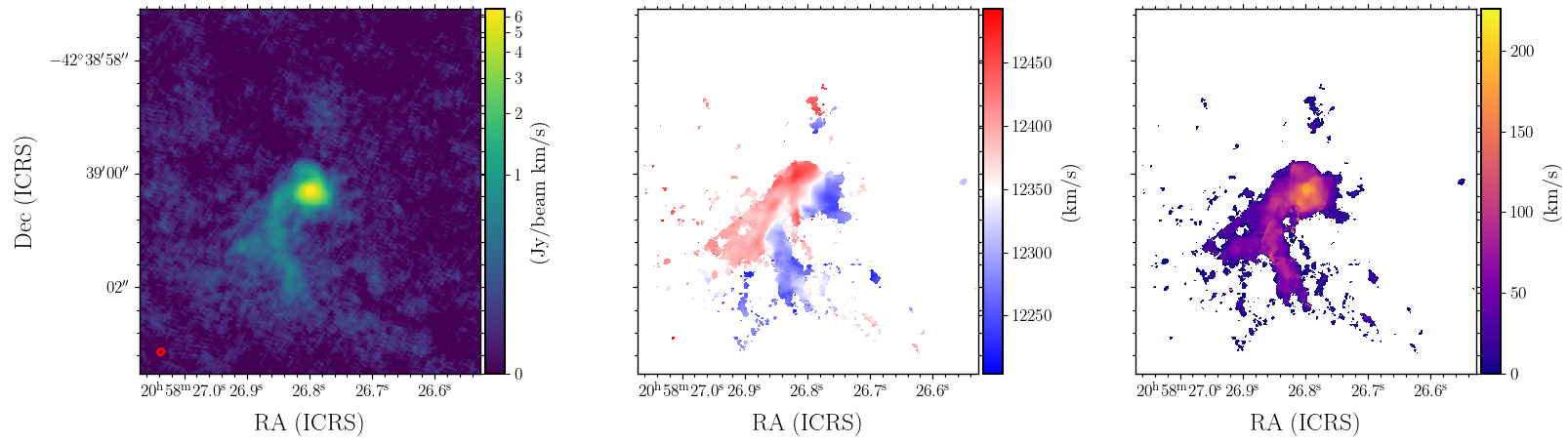}
\caption{Native resolution ALMA CO(2-1) maps of IR20551: Moment 0 (left panel), Moment 1 (middle panel) and Moment 2 (right panel).}
\label{fig:co21_maps_native_res}
\end{figure*}

\begin{figure*}
\centering
\includegraphics[width=0.4\linewidth]{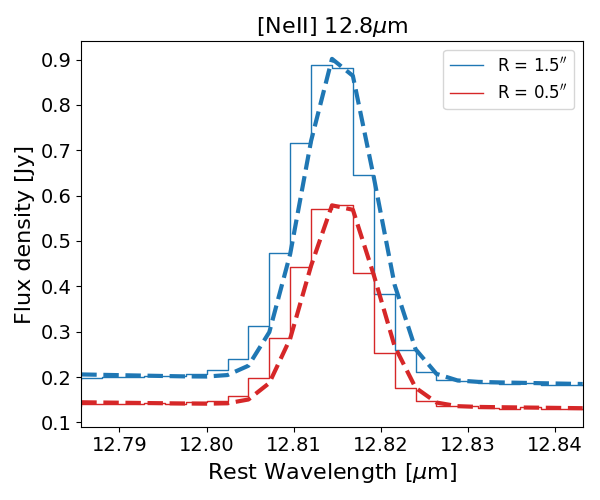}
\includegraphics[width=0.4\linewidth]{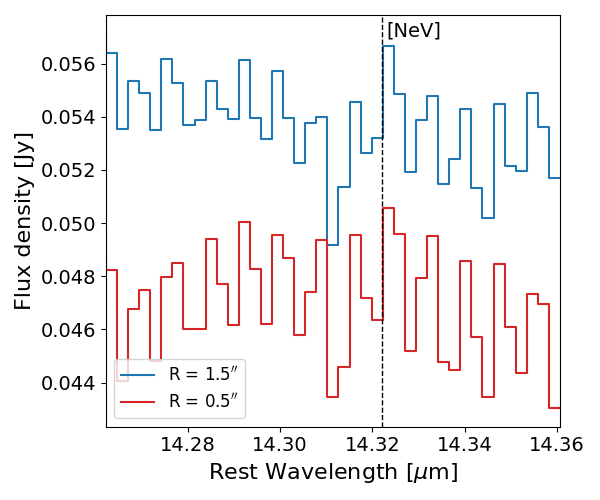}
\caption{[\ion{Ne}{ii}]12.8 (left) and [\ion{Ne}{v}]14.3 (right) emission in the nuclear and integrated MRS apertures of IR20551. [\ion{Ne}{v}] emission is absent or weak, the latter indicated by a small positive flux near its expected location.}
\label{fig:Ne_specs}
\end{figure*}

\begin{figure*}
\centering
\includegraphics[width=0.9\linewidth]{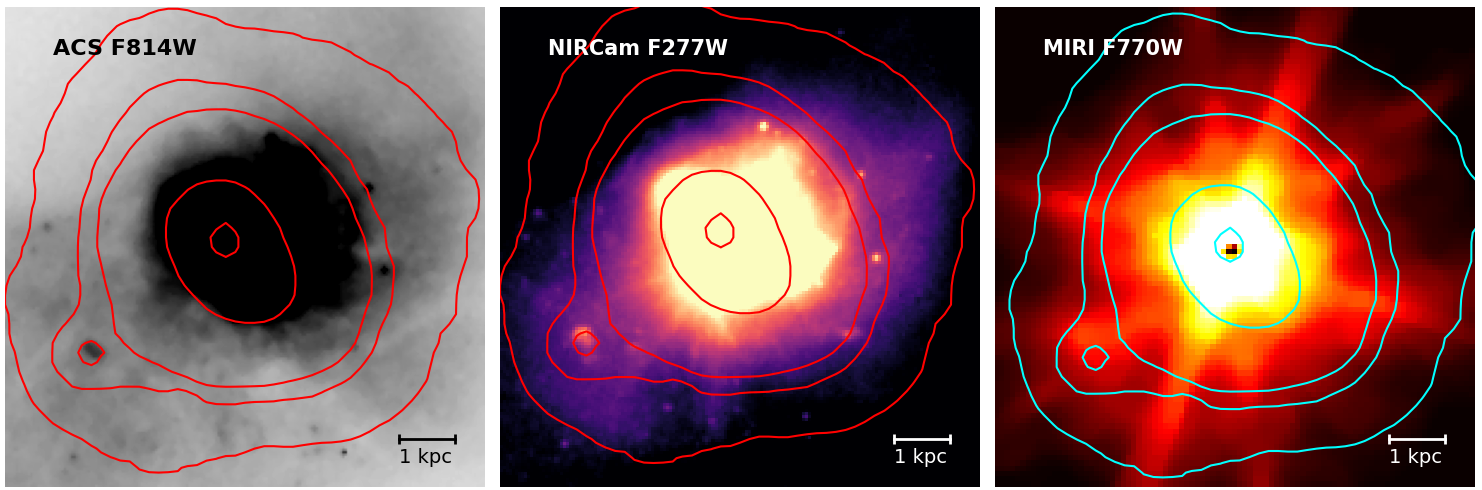}
\caption{HST ACS/F814W (left), JWST NIRCam/F277W (middle) and JWST MIRI/F770W (right) images with H$\alpha$ contours from Fig. \ref{fig:muse_fluxmaps} overlaid in red and cyan at levels 2\%, 5\%, 10\% 40\% and 95\% of the peak H$\alpha$ flux. The contours levels are chosen to visually guide the structures in the H$\alpha$ map. The second H$\alpha$ clump towards the SE of the central source clearly coincides with peaks in the F814W and F277W images.}
\label{fig:HST_NIRCam_MIRI_Halpha}
\end{figure*}

%\begin{figure*}
%\centering
%\includegraphics[width=0.9\linewidth]{PAH6.2.png}
%\caption{TBC: PAH 6.2 um map - derived via a single Gaussian fit to PAH feature at 6.2 um. PAH emission does not seem to show the same structure as the rotational transitions: e.g., the tail is absent. PAH emission appears more symmetric. If this is contribution from starburst, it validates the MUSE BPT map, which shows contribution from star forming regions in the centre. Note that this is an approximate map using a single Gaussian fit.}
%\label{fig:PAH}
%\end{figure*}

%%%%%%%%%%%%%%%%%%%%%%%%%%%%%%%%%%%%%%%%%%%%%%%%%%

% Don't change these lines
\bsp	% typesetting comment
\label{lastpage}
\end{document}